\documentclass[12pt,preprint]{emulateapj}
\shorttitle{Bow Shock at Nearby Star $\delta$ Velorum}
\shortauthors{G\'asp\'ar et al.}

\begin{document}

\title{Modeling the Infrared Bow Shock at $\delta$ Velorum: Implications for Studies 
	of Debris Disks and $\lambda$ Bo\"otis Stars}

\author{A.\ G\'asp\'ar, K.\ Y.\ L.\ Su, G.\ H.\ Rieke, Z.\ Balog}
\affil{Steward Observatory, University of Arizona, Tucson, AZ 85721}
\email{agaspar@as.arizona.edu}
\author{I.\ Kamp}
\affil{Space Telescope Science Division of ESA, STScI, 3700 San Martin Drive, 
	Baltimore, MD 21218}
\author{J.\ R.\ Mart\'inez-Galarza}
\affil{Leiden Observatory, Universiteit Leiden, 2300 RA Leiden, The Netherlands}
\author{K.\ Stapelfeldt}
\affil{Jet Propulsion Laboratory, California Institute of Technology, Pasadena, 
	CA 91109}

\begin{abstract}

We have discovered a bow shock shaped mid-infrared excess region in front
of $\delta$ Velorum using $24 ~\micron$ observations obtained with the Multiband
Imaging Photometer for {\it Spitzer} (MIPS). The excess has been classified as a 
debris disk from previous infrared observations. Although the bow shock morphology was only 
detected in the $24 ~\micron$ observations, its excess was also resolved
at $70 ~\micron$. We show that the stellar heating of 
an ambient interstellar medium (ISM) cloud can produce the measured flux. 
Since $\delta$ Velorum was classified as a debris disk star previously,
our discovery may call into question the same classification of other stars. We model 
the interaction of the star and ISM, producing images that show the same geometry and surface 
brightness as is observed. The modeled ISM is $\sim 15$ times overdense relative
to the average Local Bubble value, which is surprising considering the close proximity 
($24 ~{\rm pc}$) of $\delta$ Velorum.

The abundance anomalies of $\lambda$ Bo\"otis stars have been previously explained 
as arising from the same type of interaction of stars with the ISM. 
Low resolution optical 
spectra of $\delta$ Velorum show that it does not belong to this stellar class. The 
star therefore is an interesting testbed for the ISM accretion theory of the $\lambda$ 
Bo\"otis phenomenon. 

\end{abstract}


\section{Introduction}

Using {\it IRAS} data, more than a hundred main-sequence stars have been 
found to have excess emission in the $12$ - $100 ~\micron$ spectral range 
\citep{beckman93}. Many additional examples have been discovered with {\it ISO} and
{\it Spitzer}. In most cases the spectral energy distributions (SEDs) can be fitted
by models of circumstellar debris systems of thermally radiating dust grains with
temperatures of $50$ to $200~{\rm K}$. Such grains have short lifetimes around stars:
they either get ground down into tiny dust particles that are then
ejected by radiation pressure, or if their number density is low they are brought into 
the star by Poynting-Robertson drag. Since excesses are observed around stars that are
much older than the time scale for these clearing mechanisms, it is
necessary that the dust be replenished through collisions between
planetesimals and the resulting collisional cascades of the products of
these events both with themselves and with other bodies. Thus, planetary
debris disks are a means to study processes occurring in hundreds of
neighboring planetary sys\-tems. {\it Spitzer} observations are revealing a general
resemblance in evolutionary time scales and other properties to the events
hypothesized to have occurred in the early Solar System.

Although the planetary debris disk hypothesis appears to account for a large
majority of the far infrared excesses around main-sequence stars, there are
two alternative possibilities. The first is that very hot gas around the
stars is responsible for free-free emission \citep[e.g.,][]{cote87,su06}.
The second possibility is that the excesses arise thro\-ugh heating of
dust grains in the interstellar medium around the star, but not
in a bound structure such as a debris disk. \cite{kalas02} noticed
optical reflection nebulosities around a number of stars with Vega-like excesses.
These nebulosities show asymmetries that would not be typical of disks, they have
complex, often striated structures that are reminiscent of the Pleiades 
reflection nebulosities, and they are much too large in 
extent to be gravitationally bound to the stars \citep[see][]{gorlova06}.

\begin{deluxetable*}{cccccccc}
\tablecolumns{8}
\tablewidth{0pt}
\tablecaption{The parameters of $\delta$ Velorum \label{tab:par}}
\tablehead{
\colhead{$F_{24}$\tablenotemark{$\ast$}} & \colhead{$F_{70}$\tablenotemark{$\ast$}} & 
\colhead{$\rho_{\rm ISM}$} & \colhead{$v_{\rm rel}$} & \colhead{$F_{{\rm star} 24}$\tablenotemark{$\dagger$}}& 
\colhead{$F_{{\rm star} 70}$\tablenotemark{$\dagger$}} & \colhead{$F_{{\rm excess} 24}$\tablenotemark{$\ddagger$}}& \colhead{$F_{{\rm excess} 70}$\tablenotemark{$\ddagger$}}\\
\colhead{(mJy)} & \colhead{(mJy)} & \colhead{($10^{-24}$g cm$^{-3}$)} & \colhead{(km s$^{-1}$)} & 
\colhead{(mJy)} & \colhead{(mJy)} & \colhead{(mJy)} & \colhead{(mJy)}}
\startdata
$1569\pm42$ & $237\pm50$ & $5.8\pm0.4$ & $36\pm4$ & $1277$ & $147$ & $174$ & $141$ \\
\enddata
\tablenotetext{$\ast$}{Observed fluxes with the large aperture}
\tablenotetext{$\dagger$}{Photospheric values - not including G star component}
\tablenotetext{$\ddagger$}{Modeled excesses at large aperture}
\end{deluxetable*}

Dynamical rather than stationary interactions with the ISM are more interesting 
\citep{char91}. Originally, it was proposed that ISM dust grains could interact 
directly with material in debris disks \citep{lissauer89,whitmire92}. 
However, it was soon realized that photon pressure from
the star would repel interstellar grains, resulting in grain-free zones with
possible bow-shock geometry around luminous stars \citep{arty97}. 

This scenario has been proposed to account for the abundance anomalies
associated with $\lambda$ Bo\"otis stars. These are late B to early F-type, 
Population I stars with surface underabundances of most Fe-peak elements 
and solar abundances of lighter elements, such as C, N, O and S.
In the diffusion/accretion model \citep{venn90,kamp02,paunzen03}, 
it is suggested that the abundance anomaly occurs when
a star passes through a diffuse interstellar cloud. The radiation pressure
repels the grains, and hence much of the general ISM metals, while the gas
is accreted onto the stellar surface. While the star is within the cloud, a
mid-infrared excess will result from the heating of the interstellar dust;
however, after the star has left the cloud the abundance anomalies may
persist for $\sim 10^6 ~{\rm yr}$ in its surface layers  \citep{turcotte93} 
without an accompanying infrared excess.

There have been few opportunities to test the predictions for dynamical
interactions of main-sequence stars with the ambient interstellar
medium. \cite{france07} have studied a bow shock
generated by the O$9.5$ runaway star HD $34078$. \cite{ueta06} describe the bow
shock between the mass loss wind of the AGB star R Hya and the ISM. \cite{noriega97}
identified $58$ runaway OB stars with an observable bow shock structure
using high resolution {\it IRAS} $60 ~\micron$ emission maps. 
\cite{rebull07} discovered that the young B5 star HD 281159 is interacting with the 
ISM, producing spherical shells of extended IR emission centered on the star with a 
spike feature pointing from the star into the shells.
None of these cases correspond to the type of situation that might be mistaken
for a debris disk, nor which would be expected to produce a $\lambda$ Bo\"otis
abundance pattern. 

$\delta$ Velorum is a nearby ($\sim 24~{\rm pc}$) stellar system (at least five 
members)\footnote{It is a complex multiple system: \cite{otero00,hanbury74,horch00,
argyle02,tango79, kellerer07}}, with modest excess in the {\it IRAS} data. It has 
been classified as an A-type star with a debris disk system 
\citep[e.g.,][]{aumann85,aumann88,cote87,chen06,su06}. \cite{otero00} observed a 
drop in the primary component's brightness ($\sim 0\fm3$) and showed that it is an 
eclipsing binary with probably two A spectral type components. With the available 
data, \cite{argyle02} computed the system's parameters. They suggested that the 
eclipsing binary (Aa) consists of two A dwarfs with spectral types A1V and A5V and 
masses of $2.7$ and $2.0 ~{\rm M}_{\odot}$ and with separation of $10$ mas. 
The nearby B component is a G dwarf with mass around $1 ~{\rm M}_{\odot}$ and 
separation of $0\farcs6$ from the main component. There is also another binary 
(CD component) at $78''$ from the star.

In \S 2, we report measurements demonstrating that this star is producing a bow 
shock as it moves through an interstellar cloud as hypothesized by \cite{arty97}. 
In \S 3, we model this behavior using simple dust
grain parameters and show satisfactory agreement with expectations for the
ISM and properties of the star. We discuss these results in \S 4, where
we show that the star is most likely not part of the $\lambda$ Bo\"otis stellar class. 
Thus, $\delta$ Velorum provides a test of the diffusion/accretion hypothesis for 
$\lambda$ Bo\"otis behavior.

\section{Observations and Data Reduction}

We present observations of $\delta$ Velorum at $24$ and $70 ~\micron$ 
obtained with the Multiband Imaging Photometer for {\it Spitzer} (MIPS) as part of three 
programs: PID 57 (2004 Feb 21), PID 20296 (2006 Feb 22, Apr 3) and PID 30566 (2006 
June 12). For PID 57, we used $3$ second exposures at four dither positions with
a total integration time of 193 seconds.
The other observations at $24 ~\micron$ (PID 20296) were done in standard 
photometry mode with 4 cycles at 5 sub-pixel-offset cluster positions and 3 sec 
integrations, resulting in a total integration of 902 sec on source for each of the 
two epochs. The star HD 217382 was observed as a PSF standard (AOR ID 6627584) for PID 
57, with the same observational parameters. The observation at $70 ~\micron$ 
(PID 30566) was done in standard photometry default-scale mode with 10 sec 
integrations and 3 cycles, resulting in a total integration of 335 sec on source. 

The binary component Aa was not in eclipsing phase according to the eph\-e\-me\-ris 
equations by \cite{otero00} at either epoch. The period of the eclipse is 
$\sim 45.16 ~ {\rm days}$, and the system was $\sim 13 ~ {\rm days}$ before a
primary minimum at the first, $\sim 3 ~ {\rm days}$ before one at the second and 
$\sim 7.7 ~ {\rm days}$ before one at the third epoch for the $24 ~\micron$ 
observations. The $70 ~\micron$ observation was $2.53 ~ {\rm days}$ before a secondary 
minimum.

The data were processed using the MIPS instrument team Data Analysis
Tool \citep[DAT,][]{gordon05} as described by \cite{engelbracht07} and \cite{gordon07}.
Care was taken to minimize instrumental artifacts 
(details will be discussed in an upcoming paper, Su et al.\ 2007, in preparation).

\begin{figure*}[ht]
\figurenum{1}
\includegraphics[angle=0,scale=0.4868]{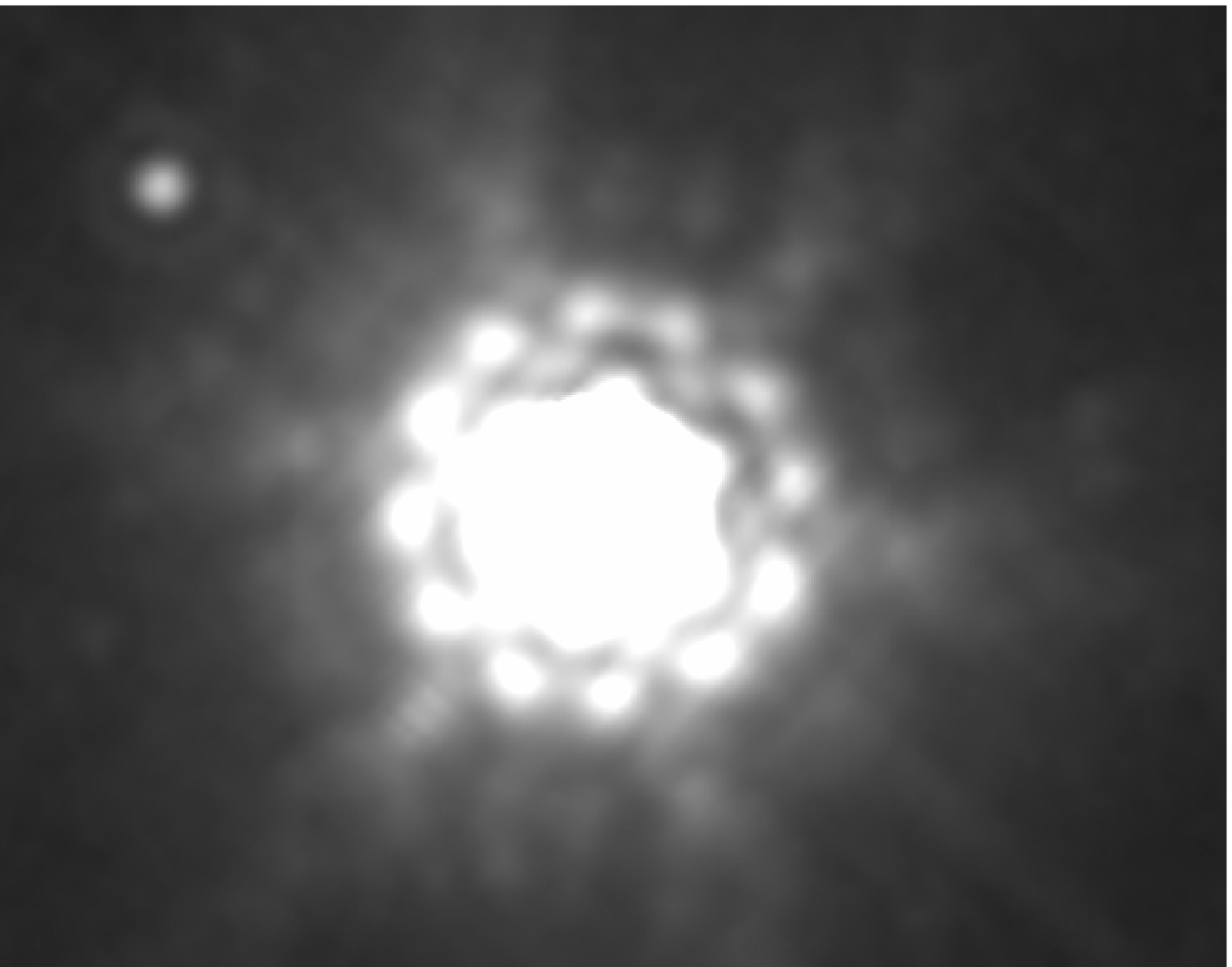} 
\includegraphics[angle=0,scale=0.4868]{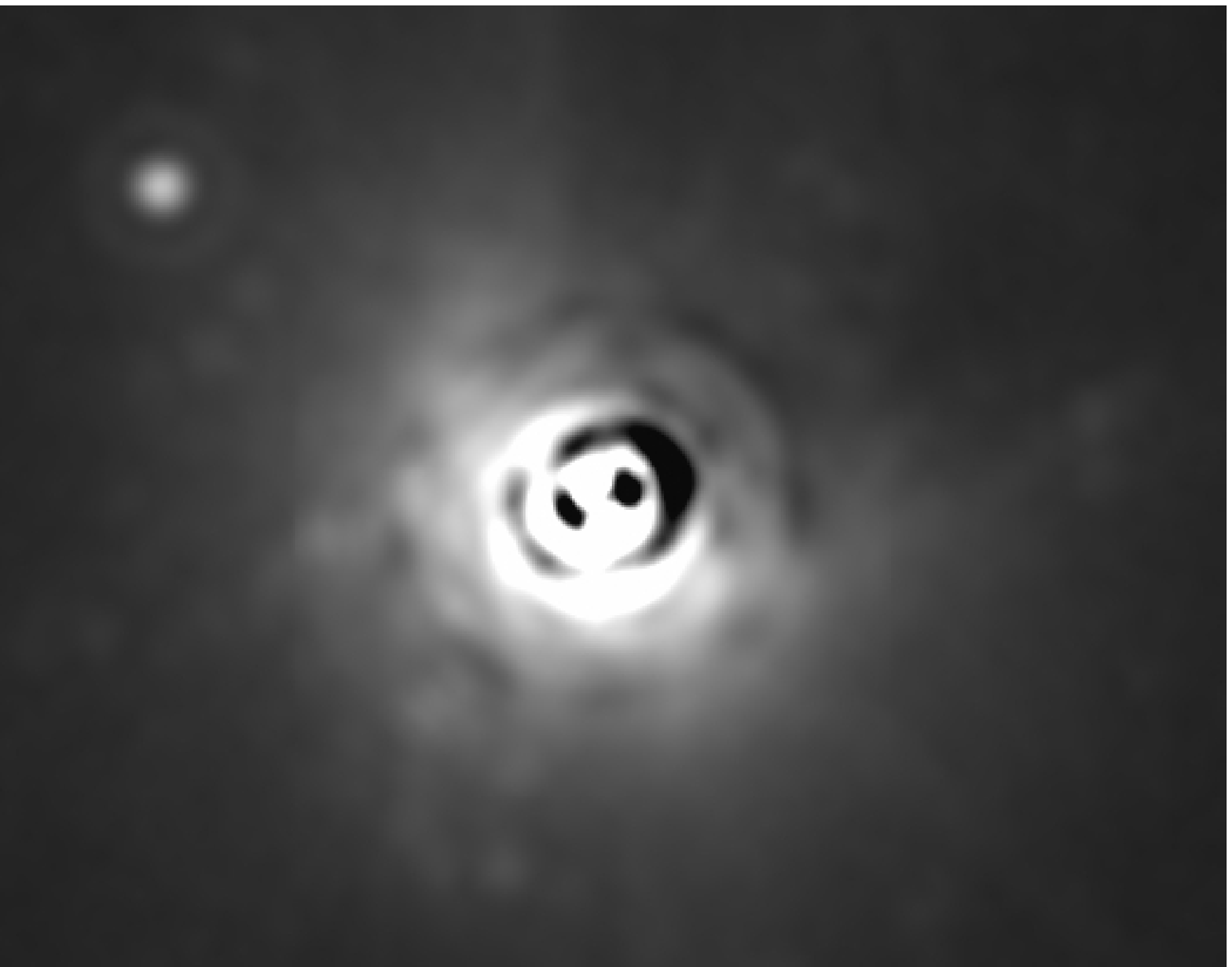} 

\includegraphics[angle=0,scale=0.4646]{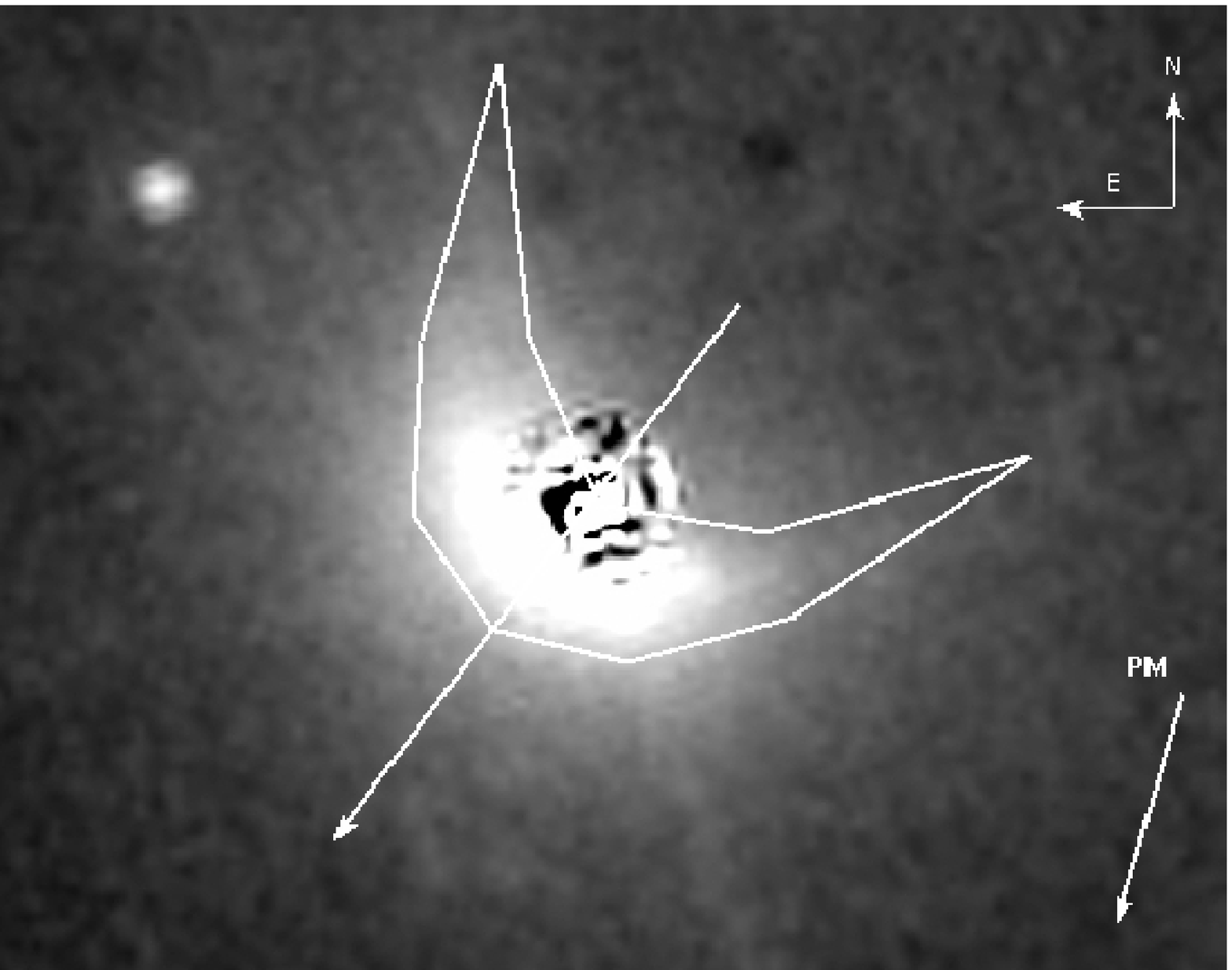} 
\includegraphics[angle=0,scale=0.4868]{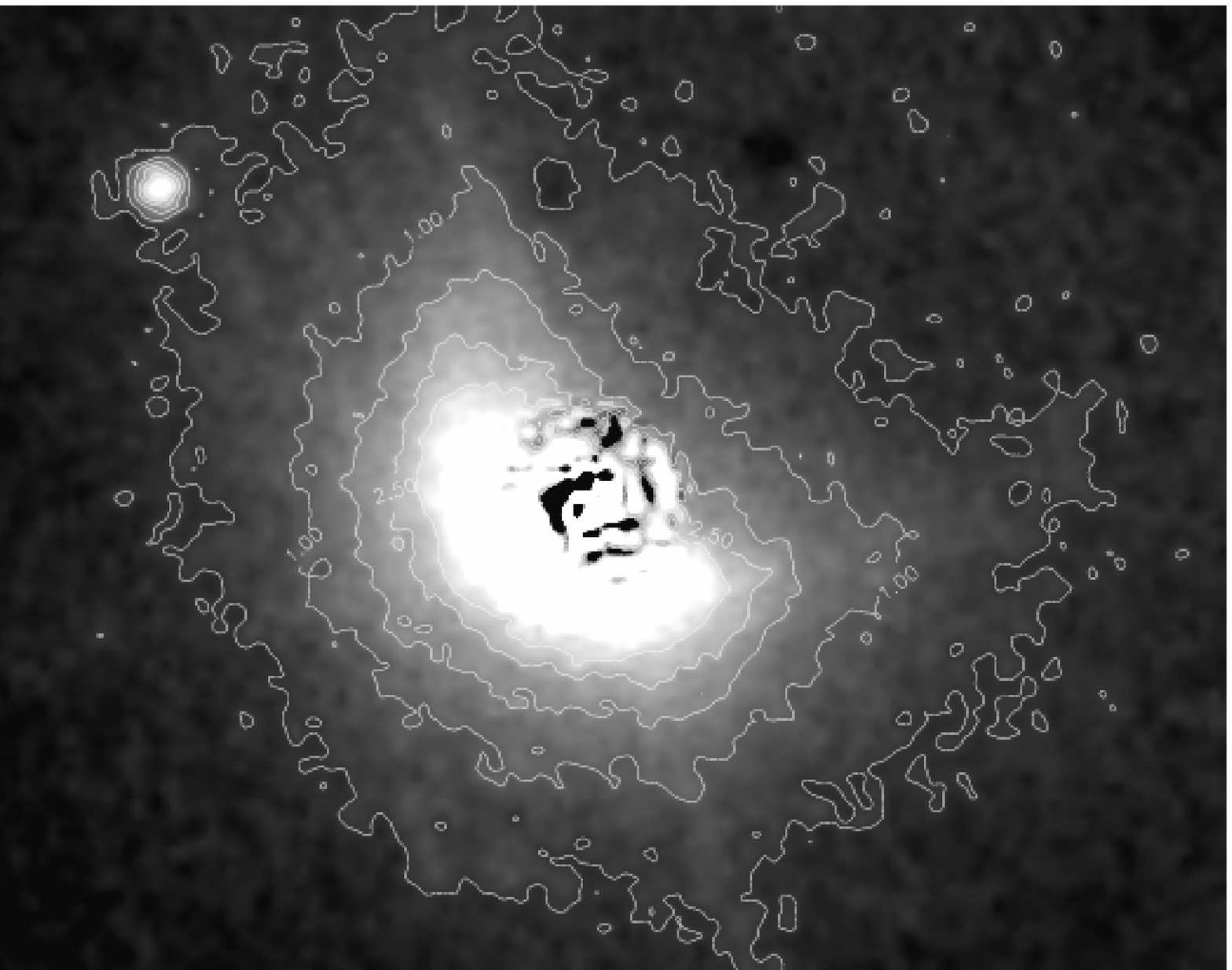} 
\caption{The panels show $24 ~\micron$ images of $\delta$ Velorum. All images are in logarithmic scaling, 
the FOV is $\sim 2\farcm74 \times 2\farcm34$. The scaling of the images are: $-0.5$ -- $4 ~{\rm MJy~sr}^{-1}$.
\textit{Top-left panel:} The original observed composite image from the $2^{\rm nd}$ and $3^{\rm rd}$ epochs. 
\textit{Top-right panel:} PSF oversubtracted image, which shows
the bow shock structure far from the star.
\textit{Bottom-left panel:} The intensity scaled PSF subtracted image (first epoch), which shows the bow shock structure close
to the star. This image shows the orientation of the images and the proper motion direction of the star. 
The arrow bisecting the bow shock contour shows the calculated direction of the modeled relative velocity.
\textit{Bottom-right panel:} Same image as the bottom-left panel, but with intensity contours plotted. The intensity 
contours are at $0.25$, $1.0$, $1.75$, $2.5$ and $3.25 ~{\rm MJy~sr}^{-1}$ from the faintest to the brightest, respectively. The contours
show that the extended emission consists of incomplete spherical shells, centered on $\delta$ Velorum.}
\label{fig:im}
\end{figure*}

Fitting the model described later demands flux 
measurements within a constant large external radius (see details in \S 3). Therefore, 
photometry for the target was extracted using aperture photometry with a single 
aperture setting. The center for the aperture 
photometry at both $24$ and $70 ~\micron$ was determined by fitting and centroiding a 
2-D Gaussian core. A radius of $56\farcs025$ was used for both wavelengths, with sky 
annulus between $68\farcs95$ and $76\farcs34$. The aperture size was chosen to be large 
enough to contain most of the flux from the bow shock, but small enough to exclude the CD 
component to avoid contamination. The CD component was bright at $24 ~\micron$ at a 
distance of $78''$ from the AaB components, but could not be detected at $70 ~\micron$. 
Aperture corrections were not applied because of the large size of the aperture. 
Conversion factors of $1.068\times10^{-3}$ and 
$1.652\times10^1 ~{\rm mJy}~{\rm arcsec}^{-2}~{\rm MIPS\_UNIT}^{-1}$ were used to transfer 
measured instrumental units to physical units at $24$ and $70 ~\micron$, respectively.

Faint extended asymmetric nebulosity offset from the central star is apparent at 
$24 ~\micron$, with the dark Airy rings partially filled in. Using standard aperture 
and point-spread-function (PSF) fitting photometry optimized for a point source, the 
total flux is $1420\pm42 ~{\rm mJy},$ $\sim 1.12$ times the expected photospheric 
flux, which was determined by fitting a Kurucz model \citep{kurucz93} 
to the optical and near infrared photometry and extrapolating it to $24$ and $70 ~\micron$.
The large aperture photometry value is greater by another factor 
of $\sim 1.1$, which puts it above the expected photospheric flux by a factor of $\sim 1.25$.
The final photometry measurements (using the large aperture setting) are listed
in Table \ref{tab:par}. We also list the modeled photospheric flux of the star and
the modeled value of the IR excess. Since the measured excess depends on the aperture
used, to avoid confusion we do not give a measured excess value, only the photospheric flux
which can be subtracted from any later measurements. The photospheric flux given in
Table \ref{tab:par} does not include the contribution from the G dwarf ($90$ and 
$10 ~{\rm mJy}$ at $24$ and $70 ~\micron$, respectively). The top left panel in 
Figure 1 shows the summed image from epochs 2 and 3, to demonstrate the asymmetry suggested 
even before PSF subtraction. 

For the first epoch $24 ~\micron$ image, the reference star image was subtracted 
from the image of $\delta$ Velorum, with a scale factor chosen as the maximum value 
that would completely remove the image core without creating significant negative flux 
residuals. The deeper exposures from the second and third epochs were designed to 
reveal faint structures far from the star, where the observed PSF is difficult to 
extract accurately. Therefore, we used simulated PSFs (from STinyTim \citep{krist02}) 
and the MIPS simulator \footnote{Software designed to simulate MIPS data, including
optical distortions, using the same observing templates used in flight.}. Because 
bright structures nearly in the PSF contribute to the residuals at large distances, 
we oversubtracted the PSF to compensate. The first epoch PSF subtracted  
$24 ~\micron$ image is shown in the bottom panels of Figure \ref{fig:im} and the
composite from epochs 2 and 3 in the upper right. 

The PSF subtracted images in Figure \ref{fig:im} show that the asymmetry is caused by 
a bow shock. As shown in the lower left, the head of the 
bow shock points approximately toward the direction of the stellar proper motion. The 
bottom right panel shows the excess flux contours and that it consists of incomplete
spherical shells centered on $\delta$ Velorum. Combined with the upper right image, there is also a 
parabolic cavity, as expected for a bow shock. The stagnation points (where photon pressure
equals gravitational force) of the grains in the bow shock are within $\sim 200 ~{\rm AU}$ 
of the star, according to the observations. A notable feature in the upper right is the 
wings of the bow shock, which are detectable to $\sim 1500 ~{\rm AU}$.

\begin{figure}[ht]
\figurenum{2}
\begin{center}
\includegraphics[angle=0,scale=0.664]{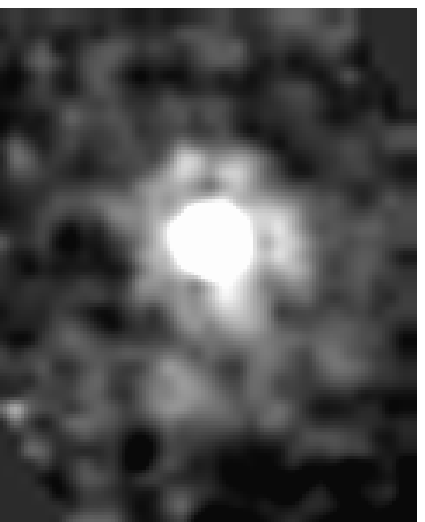}
\includegraphics[angle=0,scale=0.195]{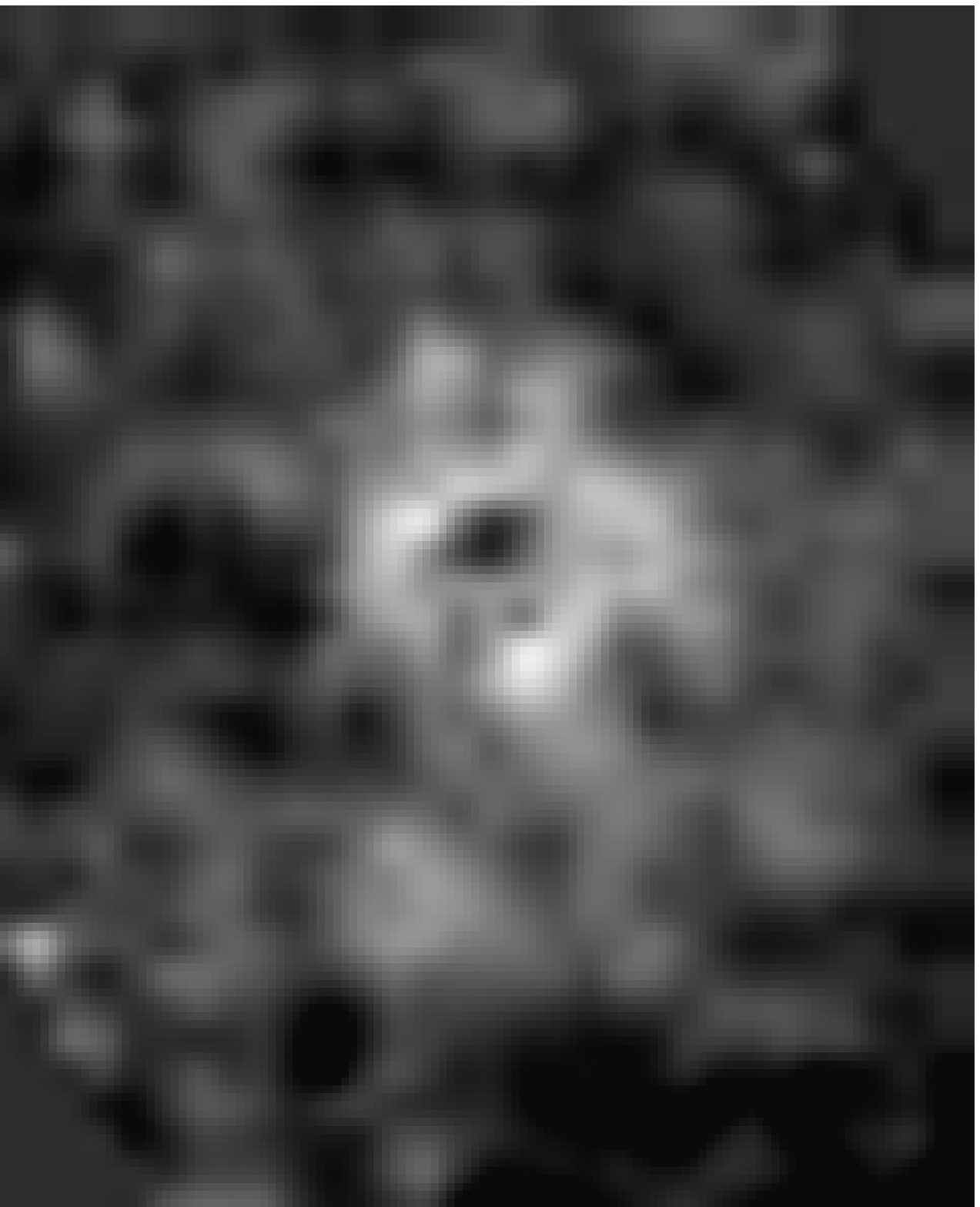}
\includegraphics[angle=0,scale=0.195]{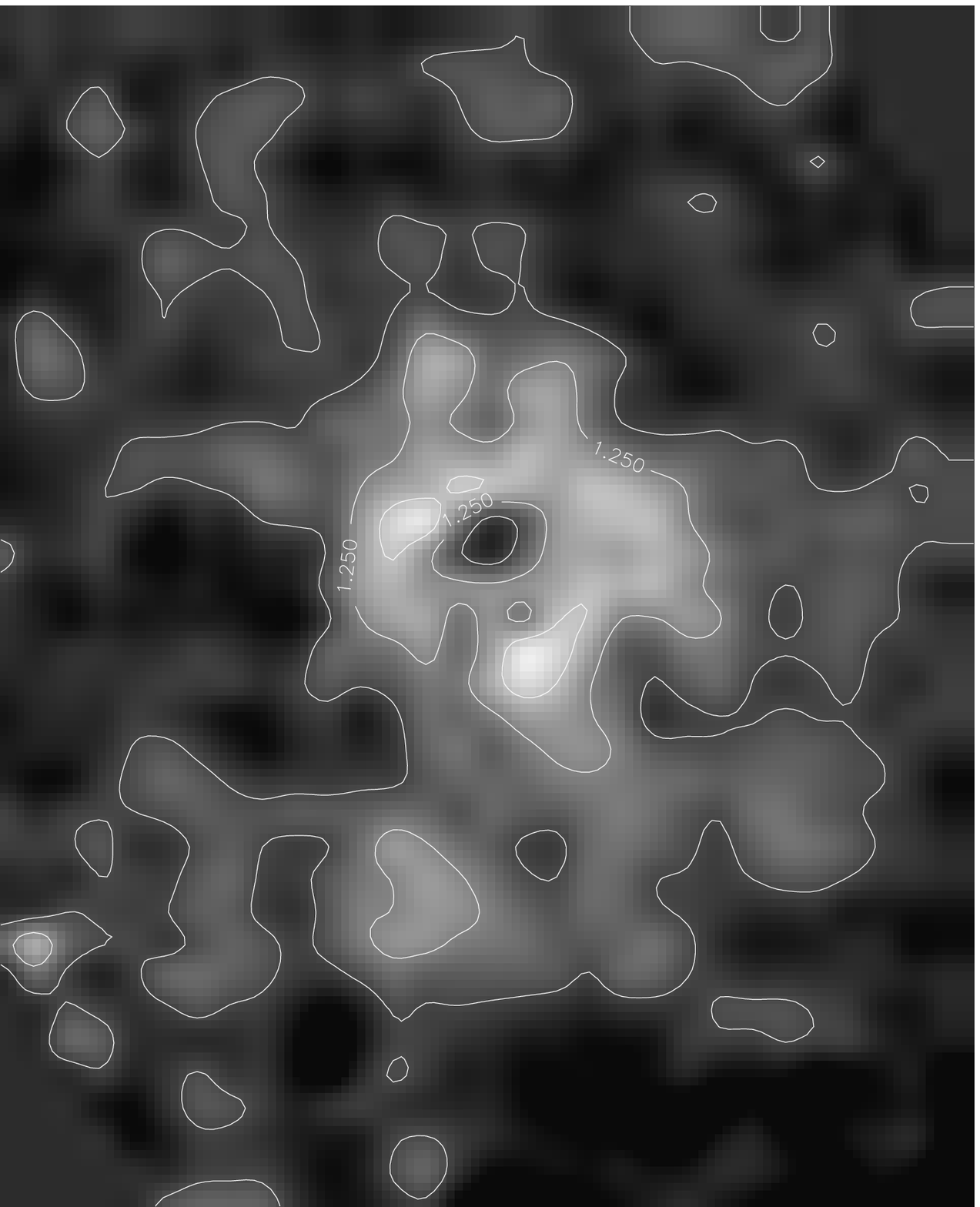}
\caption{The panels show the $70 ~\micron$ image of $\delta$ Velorum. All images are scaled logarithmically from
$-0.5$ -- $3 ~{\rm MJy~sr}^{-1}$. The FOV is $\sim 2\farcm46 \times 3\farcm03$. The orientation
of the images is the same as in Figure \ref{fig:im}. {\it First panel:}
the observed image. {\it Middle panel:} the PSF subtracted image. The residual flux seems
close to being concentric. {\it Last panel:} the intensity contours. They suggest that there is a faint concentric
$70 ~\micron$ excess further from the star that fades at the cavity region behind the star.}
\label{fig:70}
\end{center}
\end{figure}

The $70 ~\micron$ observation is shown in Figure \ref{fig:70}. The PSF subtraction 
(scaled to the point source flux of $125 ~{\rm mJy}$) does not reveal the bow shock 
structure at this wavelength, only that there is extended excess. The total flux of the 
residual of the PSF subtracted image is $119 ~{\rm mJy}$. The intensity contours ({\it 
last panel}) suggest that the $70 ~\micron$ excess fades at the cavity behind the 
star, but the effect is small. The geometry and direction of the bow shock are 
discussed in more detail in \S 3.2.

\section{The Bow Shock Model}

Based on a previous suggestion by \cite{venn90}, \cite{kamp02} proposed a 
physical model to explain the abundance pattern of $\lambda$ Bo\"otis stars through 
star-ISM interaction and the diffusion/accretion hypothesis. Their model is based on a 
luminous main-sequence star passing through a diffuse ISM cloud. The star blows the 
interstellar dust grains away by its radiation pressure, but accretes the interstellar 
gas onto its surface, thus establishing a thin surface layer with abundance anomalies. 
So long as the star is inside the cloud, the dust grains are heated to produce excess 
in the infrared above the photospheric radiation of the star. 
Mart\'inez-Galarza et al.\ (2007, in prep.) have developed a model of this process and 
show that the global spectral energy distributions of a 
group of $\lambda$ Bo\"otis type stars that have infrared excesses are consistent with 
the emission from the hypothesized ISM cloud. Details of the model can be found in 
their paper. Here we adapt their model and improve its 
fidelity (e.g., with higher resolution integrations), and also model the surface 
brightness distribution to describe the observed bow shock seen around $\delta$ Velorum.

\subsection{Physical description of the model}

The phenomenon of star-ISM interactions generating bow shocks was first 
studied by \cite{arty97}. They showed that the radiative pressure force on a 
sub-micron dust grain can be many times that of the gravitational force as it 
approaches the star. The scattering surface will be a parabola with the star at the 
focus point of the parabolic shaped dust cavity. Since the star heats the grains 
outside of the cavity and close to the parabolic surface, an infrared-emitting bow 
shock feature is expected. 

The shape of the parabola (for each grain size) can be given in terms of the distance 
between the star (focus) and the vertex. This so-called avoidance radius (or the 
$p/2$ parameter of the scattering parabola) can be calculated from energy conservation 
to be \citep{arty97}:
\begin{equation}
r_{\rm av}^a=\frac{2\left(\beta^a -1\right){\rm G}M}{v_{\rm rel}^2},
\end{equation}
where $a$ is the radius of the particle, $M$ is the mass of the star and $v_{\rm rel}$ 
is the relative velocity between the star and the dust grains. 

\begin{figure*}[ht]
\figurenum{4}
\includegraphics[angle=0,scale=0.54]{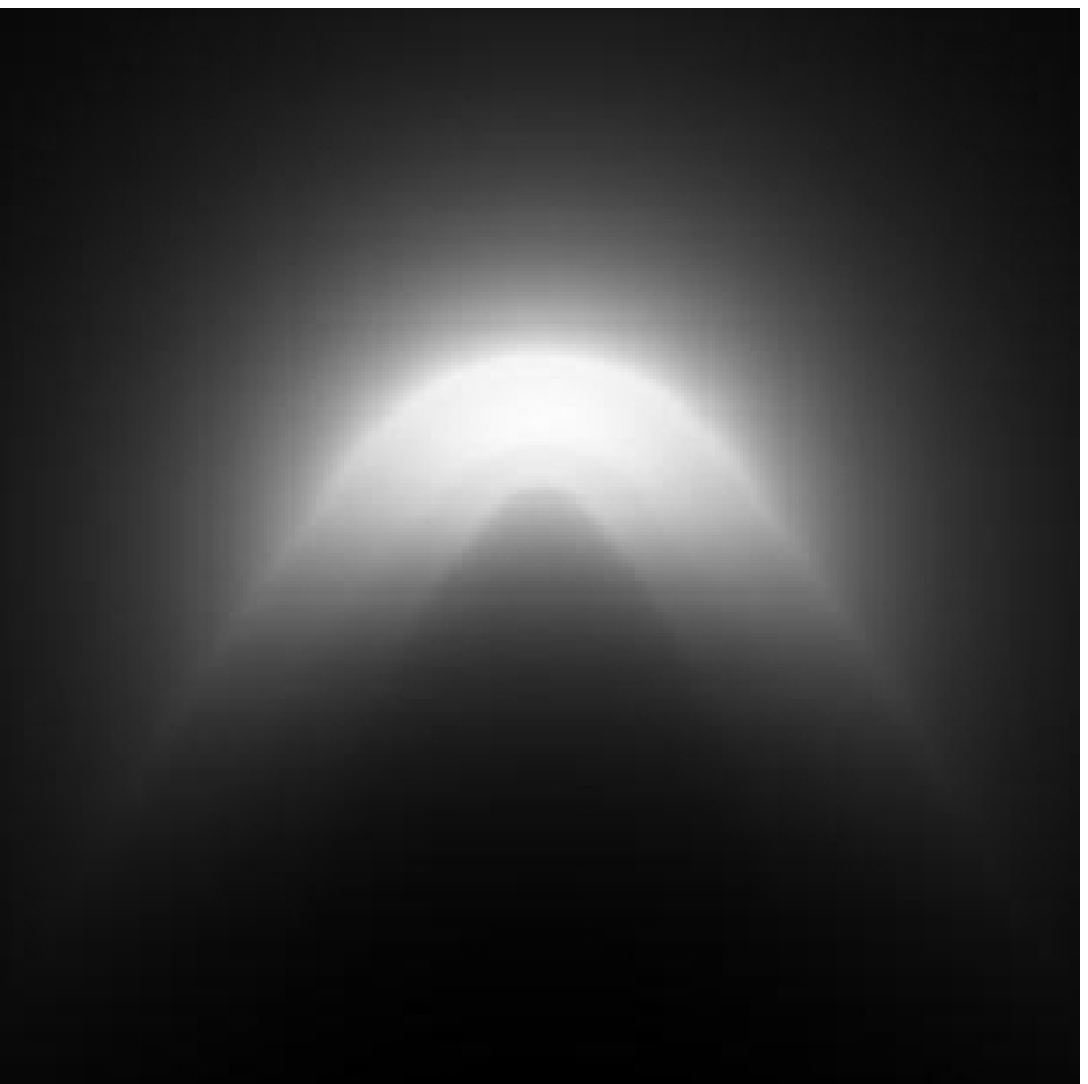}
\includegraphics[angle=0,scale=0.54]{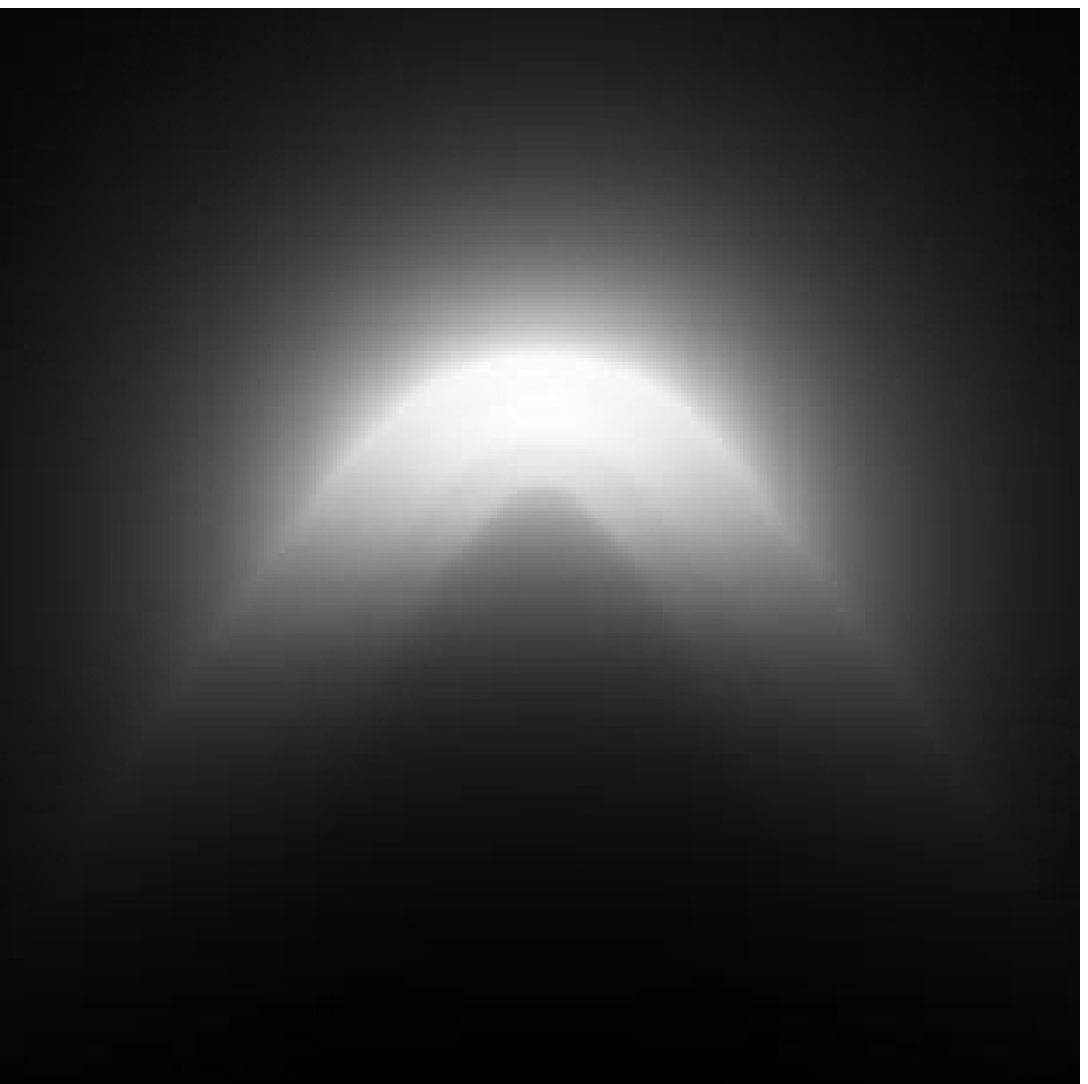}
\includegraphics[angle=0,scale=0.54]{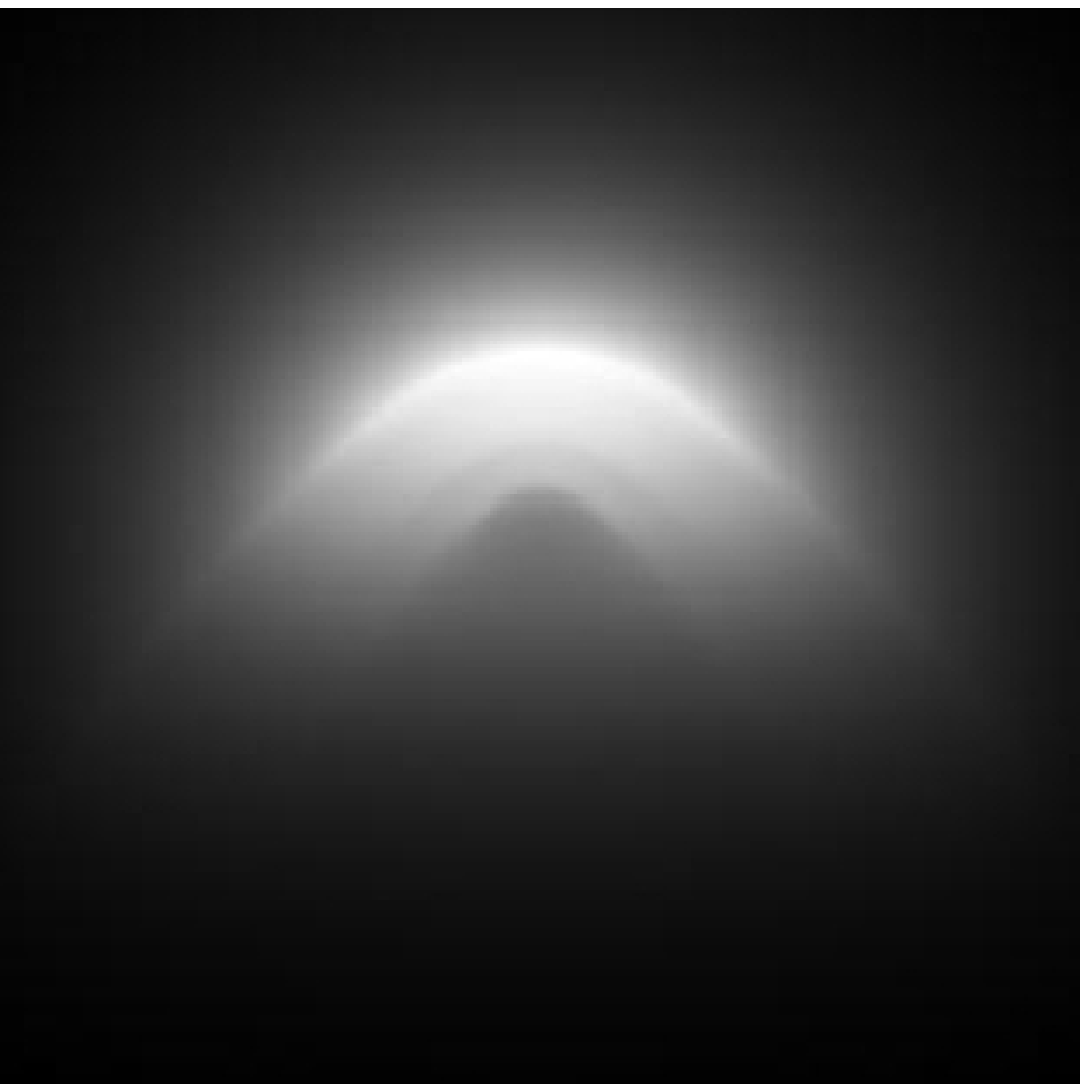}
\caption{The panels show the $24 ~\micron$ morphology of the bow shock viewed at different inclinations, starting
from $90^{\circ}$ ({\it left}), $70^{\circ}$ ({\it middle}) and $50^{\circ}$ ({\it right}).}
\label{fig:slides}
\end{figure*}

$\beta^a$ is the ratio of photon pressure to gravitational force on a grain and it is 
given by \citep{burns79}:
\begin{equation}
\beta^a=0.57 Q_{\rm pr}^a\frac{L/L_{\odot}}{M/M_{\odot}}\left(\frac{a}{\micron}\right)^{-1}\left(\frac{\delta}{{\rm g}~{\rm cm}^{-3}}\right)^{-1},
\end{equation}
where $\delta$ is the bulk density of the grain material and $Q_{\rm pr}^a$ is the 
radiation pressure efficiency averaged over the stellar spectrum. 
$Q_{\rm pr}^a(\lambda)$ can be expressed in terms of grain properties 
\citep[absorption coefficient $Q_{\rm ab}^a(\lambda)$, scattering coefficient 
$Q_{\rm sca}^a(\lambda)$ and the scattering asymmetry factor 
$g=\left<\cos\alpha(\lambda)\right>$,][]{burns79,henyey38}:
\begin{equation}
Q_{\rm pr}^a(\lambda)= Q_{\rm ab}^a(\lambda)+ Q_{\rm sca}^a\left(1-g\right),
\end{equation}
which gives
\begin{equation}
Q_{\rm pr}^a = \frac{\int{Q_{\rm pr}^a(\lambda)B(T_{\ast},\lambda)}{\rm d}\lambda}{\int{B(T_{\ast},\lambda)}{\rm d}\lambda},
\end{equation}
where $B(T_{\ast},\lambda)$ is the Planck function. We adopted astronomical silicates in 
our model with $\delta = 3.3 ~{\rm g~cm}^{-3}$ from \cite{draine84} and 
\cite{laor93}. We considered a MRN \citep{mathis77} grain size distribution in our 
model:
\begin{equation}
{\rm d}n = C a^{-\gamma}{\rm d}a,
\end{equation}
where $C$ is a scaling constant and $n$ is the number density of the cloud with 
$\gamma=3.5$ and grain sizes ranging from $0.005$ -- $0.25 ~{\rm \micron}$.

With these equations we are able to model the avoidance cavity for a grain that 
encounters a star with known mass, luminosity and relative velocity. The model 
describes a situation where the expelled grains are instantly removed from the system 
rather than drifting away, but this only causes a minor discrepancy in the wing and 
almost none in the apex of the parabola compared to the actual case. In the actual 
scenario only those particles get scattered back upstream that encounter the central 
star with small impact parameter ($\sim r_{\rm av}^a/2$). This means that most of the 
grains will get expelled toward the wings, where the grains go further out and emit less 
infrared excess, thus their contribution to the total flux will be small.

The model determines the number density of certain grain sizes and the position of their 
parabolic avoidance cavity. Outside of the cavity we assumed a constant number density 
distribution for each grain size. To calculate the surface brightness of the system and 
its SED we assumed a thermal equilibrium condition, with wavelength dependent absorption 
and an optically thin cloud. 

\subsection{Model Geometry and Parameters}

The model described in \S 3.1 gives the distribution and temperature for each 
grain size. This model was implemented in two ANSI C programs. The first program fits 
the SED of the system to the observed photometry points, while the second program 
calculates the surface brightness of the system. The fitted photometry included uvby, 
UBV, {\it HIPPARCOS} V band, 2MASS, {\it IRAS} and MIPS ($24$ and $70 ~\micron$) data.
We subtracted the $24$ and $70 ~\micron$ flux contributed by the G star 
($90$ and $10 ~{\rm mJy}$, respectively) from the MIPS observations, because we wanted 
to model the system consisting of the two A stars and the bow shock.

\begin{figure}[ht]
\figurenum{3}
\begin{center}
\includegraphics[angle=0,scale=0.55]{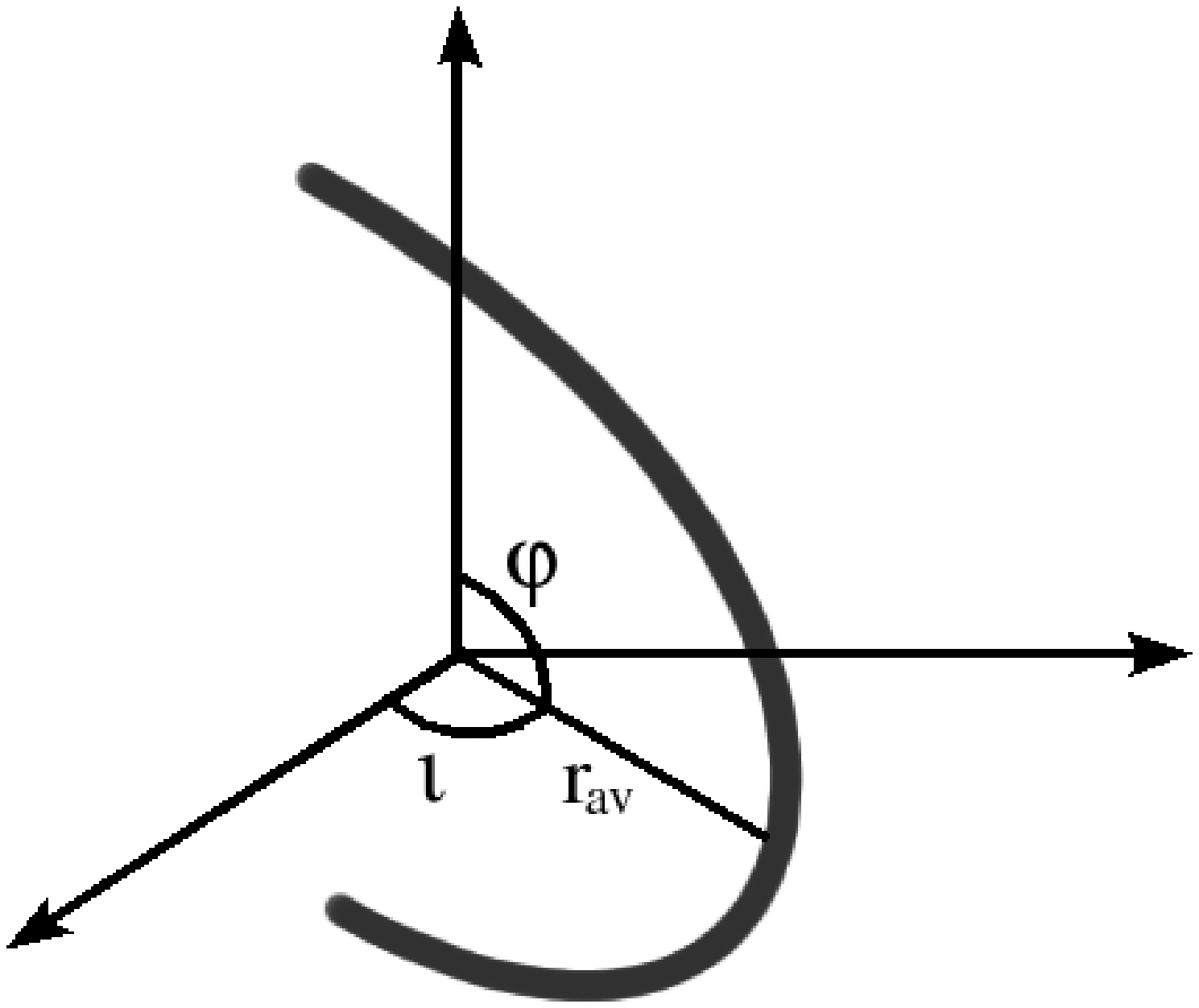}
\caption{The nomenclature of the angles of the system. The heavy line is the grain avoidance parabola. $\varphi$ is the rotation angle of the system on the plane
of the sky (our initial guess was $4^{\circ}$ N from the calculated direction of relative motion shown in 
Figure \ref{fig:im}), $\iota$ is the inclination and ${\rm r}_{\rm av}$ is an avoidance radius. The observer is viewing from 
the axis pointing to the bottom left.}
\label{fig:angles}
\end{center}
\end{figure}

\begin{figure*}[ht]
\figurenum{5}
\includegraphics[angle=0,scale=0.88]{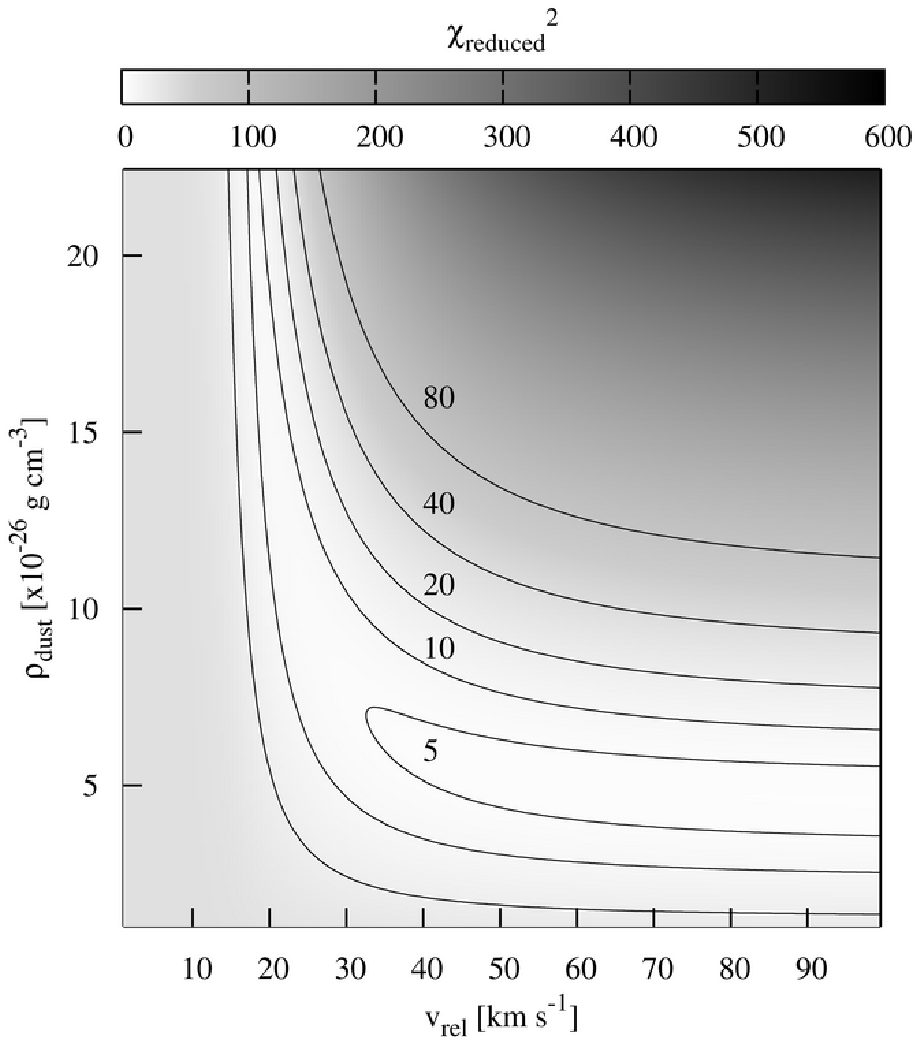} 
\includegraphics[angle=0,scale=0.895]{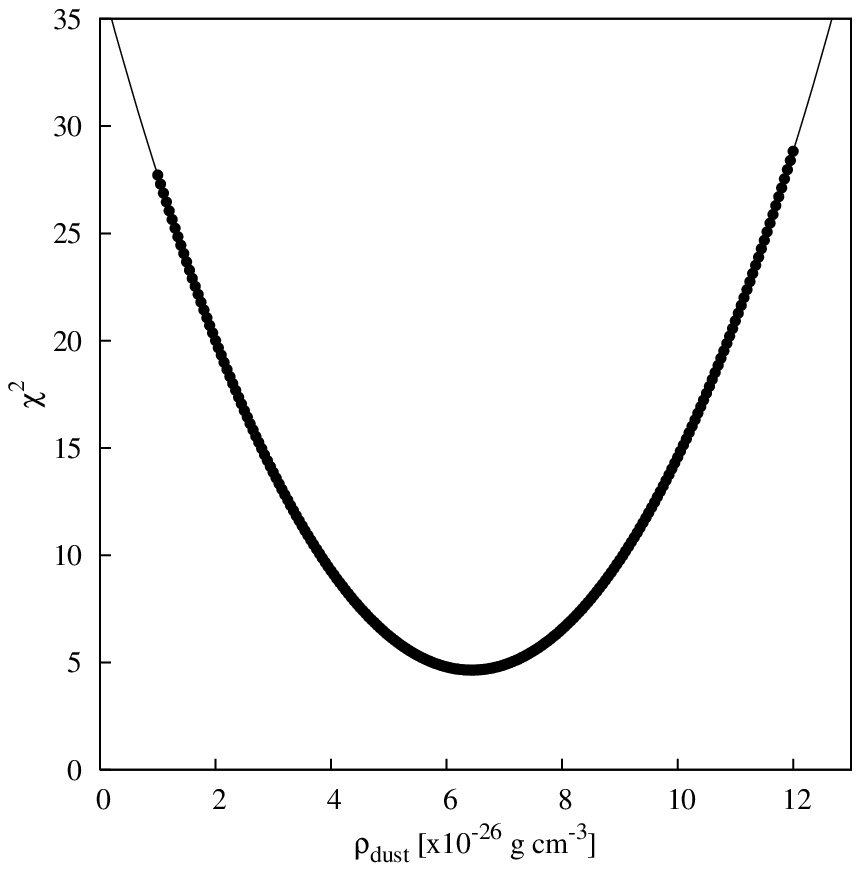} 
\caption{{\it Left panel:} The $\chi^2$ phase space for $\rho_{\rm dust}$ vs.\ ${\rm v}_{\rm rel}$ with constrained $r_{\rm ext}=1366 ~{\rm AU}$. {\it Right panel:} The $\rho$ vs.\ $v_{\rm rel}$ phase space ({\it left
panel}) cut at $v_{\rm rel} = 36 ~{\rm km~s}^{-1}$.}
\label{fig:chi}
\end{figure*}

The input parameters are: stellar radius, mass-to-luminosity ratio (MLR), relative 
velocity of cloud and star, ISM dust density, cloud external radius and the distance of 
the system. The stellar radius, MLR and the distance can be constrained easily. We
determined the best-fit Kurucz model \citep{kurucz93} by fitting the photometry points 
at wavelengths shorter than $10 ~\micron$. Since the distance is known to high accuracy 
from {\it HIPPARCOS} we can determine the radius and thus the luminosity of the star 
precisely. The mass was adopted from \cite{argyle02}. The G dwarf's luminosity is only
1\% of the system, so leaving the star out does not cause any inconsistency. Its mass is only 17\% of
the total mass, which can only cause minor changes in the determined final relative velocity,
but none in the final surface brightness or the computed ISM density. 
The model then has three variable parameters: the density of the ISM grains 
($\rho_{\rm dust}$ - does not include gas), the relative velocity between the cloud and the star 
($v_{\rm rel}$) and the external radius of the cloud ($r_{\rm ext}$). The model should describe 
the total flux from exactly the area used for our photometry. The aperture 
radius of $56\farcs025$ ($1366 ~{\rm AU}$ at the stellar distance of 
$24.45 ~{\rm pc}$) was used as $r_{\rm ext}$. Both programs 
calculate the $Q_{\rm pr}^a$, $\beta^a$, $r_{\rm av}^a$, $n$ values and then the 
temperature at $r_{\rm av}^a$ for each grain size. 

The SED modeling program decreases the temperature value from the one at $r_{\rm av}$ 
by $0.01 ~{\rm K}$ steps and finds the radius for that corresponding grain temperature. 
The program does not include geometrical parameters such as the inclination or the 
rotation angle of the system, since these are irrelevant in calculating the total flux. 
It calculates the contribution to the emitting flux for every grain size from every 
shell to an external radius ($r_{\rm ext}$) and adds them up according to wavelength. 

The program that calculates the surface brightness uses a similar algorithm as the SED 
program, but it calculates the temperature at $1 ~{\rm AU}$ distance steps from 
$r_{\rm av}^a$ for every grain size and calculates the total flux in the line of sight 
in $1 ~{\rm AU}^2$ resolution elements. 

The total inclination $\iota$ of the bow shock was not included as a parameter, since 
by eye the observed images seemed to show an inclination of $\iota \approx 90^{\circ}$ 
(a schematic plot of the angle nomenclature is shown in Figure \ref{fig:angles}). This 
approximation is strengthened by the radial velocity of the star, which is only 
$\sim 2 ~{\rm km~s}^{-1}$ compared to the tangential velocity of 
$\sim 13 ~{\rm km~s}^{-1}$. This assures that the motion of the system is close to 
perpendicular to the line of sight. However, we have found that the bow shock has similar 
appearance for a significant range of angles ($\pm 20^{\circ}$) relative to $\iota = 90^{\circ}$.
We illustrate this in Figure \ref{fig:slides}. If the relative velocity vector 
would have a $70^{\circ}$ (or $110^{\circ}$) inclination it would only cause minor differences 
in the modeled velocity ($\Delta v_{\rm rel}\approx 3 ~{\rm km~s}^{-1}$) and ISM density ($\Delta
\rho_{\rm ISM}\approx 0.2\times10^{-24}~{\rm g~cm}^{-3}$). At an 
inclination of $50^{\circ}$, the ``wings'' spread out and the bright rim at the apex starts to 
become thin.

With interstellar FeII and MgII measurements \cite{lallement95} showed that the Local 
Interstellar Cloud (LIC) has a heliocentric velocity of $26 ~{\rm km~s}^{-1}$ moving 
towards the galactic coordinate $l_{II}=186\pm3^{\circ}$, $b_{II}=-16\pm3^{\circ}$. 
Since $\delta$ Velorum is at $l_{II}\approx272^{\circ}$, $b_{II}\approx-7^{\circ}$, the 
LIC is also moving perpendicular to our line of sight at the star and in the direction 
needed to reach a high relative velocity between the star and cloud. \cite{crawford98} 
showed a low velocity interstellar Ca K line component in the star's spectrum with 
$v_{\rm helio}=1.3\pm0.4~{\rm km~s}^{-1}$, which also proves that the ISM's motion is 
perpendicular to our line of sight at $\delta$ Velorum. The offset of the proper motion
direction of the star from the head direction of the bow shock by a few degrees could be 
explained by the ISM velocity. A simple vectorial summation of the star and the ISM 
velocities should give a net motion in the direction of the bow shock.

\begin{figure*}[ht]
\figurenum{7}
\includegraphics[angle=0,scale=1.165]{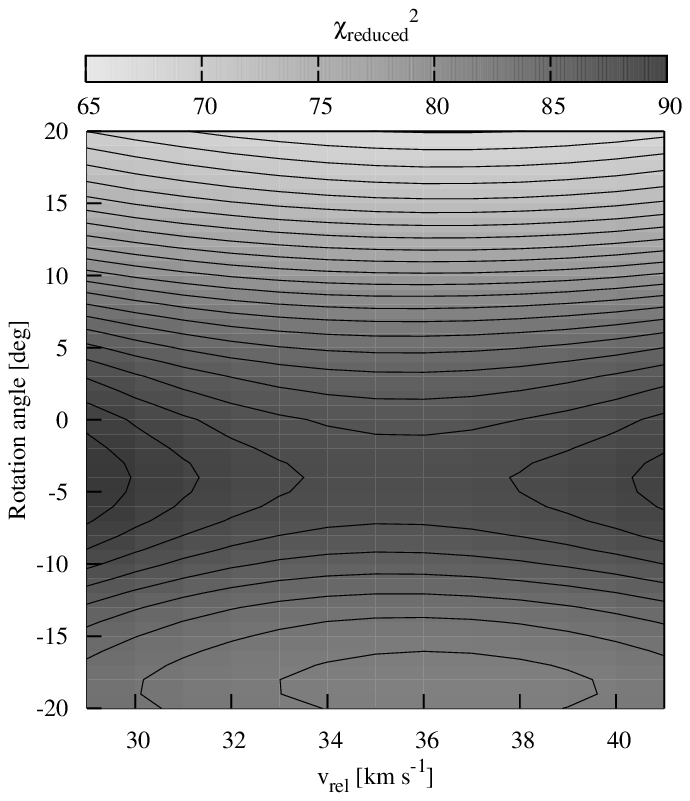} 
\includegraphics[angle=0,scale=.91]{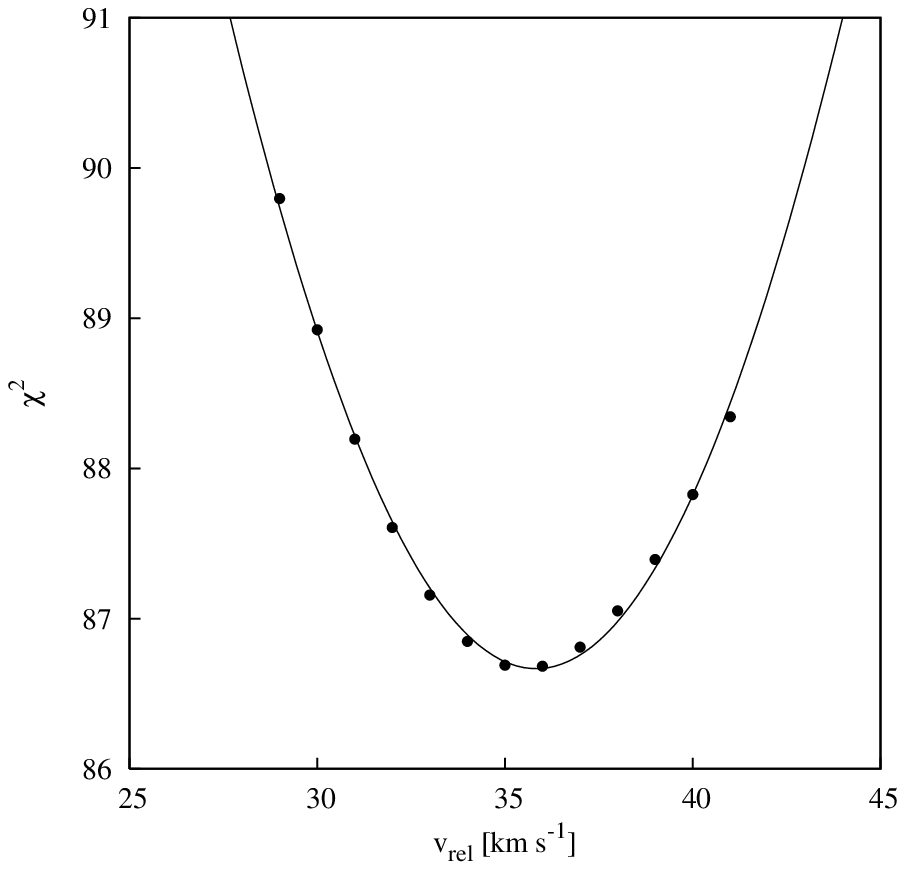}   
\caption{{\it Left panel:} $\chi^2$ in the phase space of $\varphi$ vs.\ $v_{\rm rel}$. {\it Right panel:} The phase space cut at $\varphi=-4^{\circ}$, showing the best
fit for $v_{\rm rel}$.}
\label{fig:bestrot}
\end{figure*}

\begin{figure}[ht]
\figurenum{6}
\begin{center}
\includegraphics[angle=0,scale=.68]{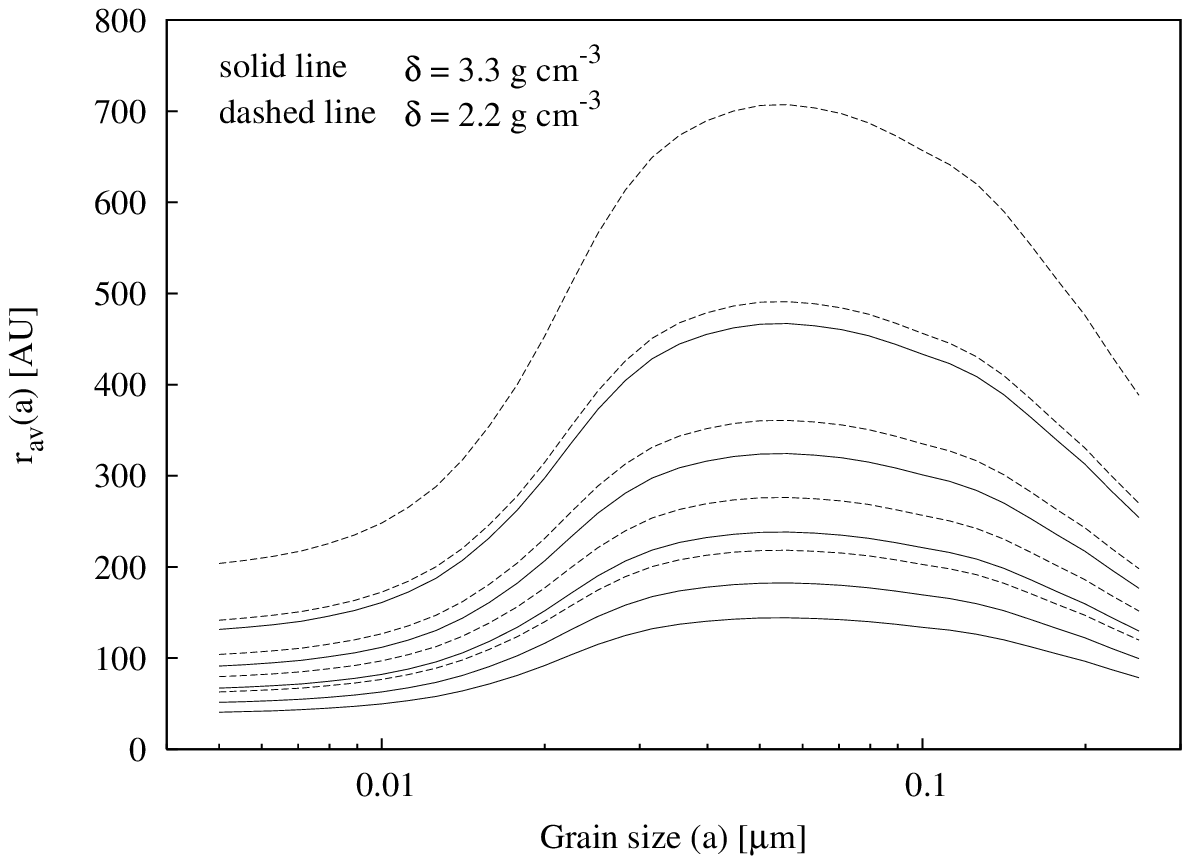}
\caption{The value of $r_{\rm av}$ as a function of grain size. The solid lines are curves for a silicate bulk density of
$3.3 ~{\rm g~cm}^{-3}$, while the dashed ones are
for $2.2 ~{\rm g~ cm}^{-3}$. The curves are for $v_{\rm rel}$ values of $25$, $30$, $35$, $40$ and $45 ~{\rm km~s}^{-1}$ from top to bottom, respectively.}
\label{fig:ravvsgr}
\end{center}
\end{figure}

\subsection{Results}
We first tried to find the best fitting SED to the photometry points 
corresponding to wavelengths larger than $10 ~\micron$ (MIPS, {\it IRAS}) with $\chi^2$ 
minimization in the $v_{\rm rel}$ vs.\ $\rho_{\rm dust}$ phase space. We defined $\chi^2$ as:
\begin{equation}
\chi^2=\sum\frac{\left({\rm F}_{\rm obs}-{\rm F}_{\rm calc}\right)^2}{\sigma_{\rm obs}^2}
\end{equation}
The $\chi^2$ phase space with $r_{\rm ext}=1366 ~{\rm AU}$ showed no minimum (Figure 
\ref{fig:chi}, {\it left panel}). The interpretation of the diagram is as follows: if 
the relative velocity is small, then the avoidance radius will be large. Consequently 
the grains will be at relatively low temperature and the amount of dust required to 
produce the observed flux increases. On the other hand, if the relative velocity is 
large, then the grains can approach closer to the star and heat up to higher temperatures. 
As a result a smaller dust density is enough to produce the observed flux. Therefore, 
the combination of the density of the cloud and the relative velocity can be well 
constrained by the broad-band SED alone, but not each separately.

By using surface brightness values from the observations and the model calculations we 
were able to determine the $v_{\rm rel}$ parameter and thus eliminate the degeneracy
of the model. Since the bow shock is a parabolic feature it has only one variable, the 
avoidance radius ($r_{\rm av}^a$), which is the same as the $p/2$ parameter of the 
parabola (with $p$ being the distance between the focus point and the vertex). The value 
of $r_{\rm av}^a$ does change as a function of grain size, but the head of the bow shock 
will be near the value where the avoidance radius has its maximum as a function of grain 
size. As can be seen in Figure \ref{fig:ravvsgr}, the avoidance radius has a maximum at 
$\sim 0.06 ~\micron$ grain size. The value of the avoidance radius on the other hand 
only depends on the relative velocity between the ISM cloud and the star. This way we 
can constrain the second parameter of the model ($v_{\rm rel}$). The relative velocity 
has to be set so that the avoidance radius of the $\sim 0.06 ~\micron$ grain is around 
half the parabola parameter value. This method gives a value that only approximates the 
true one, but it can be used as an initial guess. 

The $v_{\rm rel}$ parameter was constrained by comparing the PSF subtracted image 
``wings'' with model images. Within a range of $\pm 6 ~{\rm km~s}^{-1}$ of our initial 
guess ($v_{\rm rel}= 35 ~{\rm km~s}^{-1}$) with $1 ~{\rm km~s}^{-1}$  steps, we 
generated images of the surface brightness distribution to a radius of $2500 ~{\rm AU}$. 
The computational time for a total $5000 \times 5000 \times 5000 ~{\rm AU}$ data cube 
was long, so we only calculated to a depth of $250 ~{\rm AU}$, keeping the field of 
view (FOV) $5000 \times 5000 ~{\rm AU}$. The fluxes of the generated images were normalized
(to ensure that the geometry was the main constraint of the fit and not surface 
brightness variations) and rotated to angles $\varphi=\pm 20^{\circ}$ with $1^{\circ}$ 
steps. After rotation, both the model images and the observed image were masked with 
zeros where there was no detectable surface brightness in the observed image.

\begin{figure*}[ht]
\figurenum{11}
\includegraphics[angle=0,scale=.321]{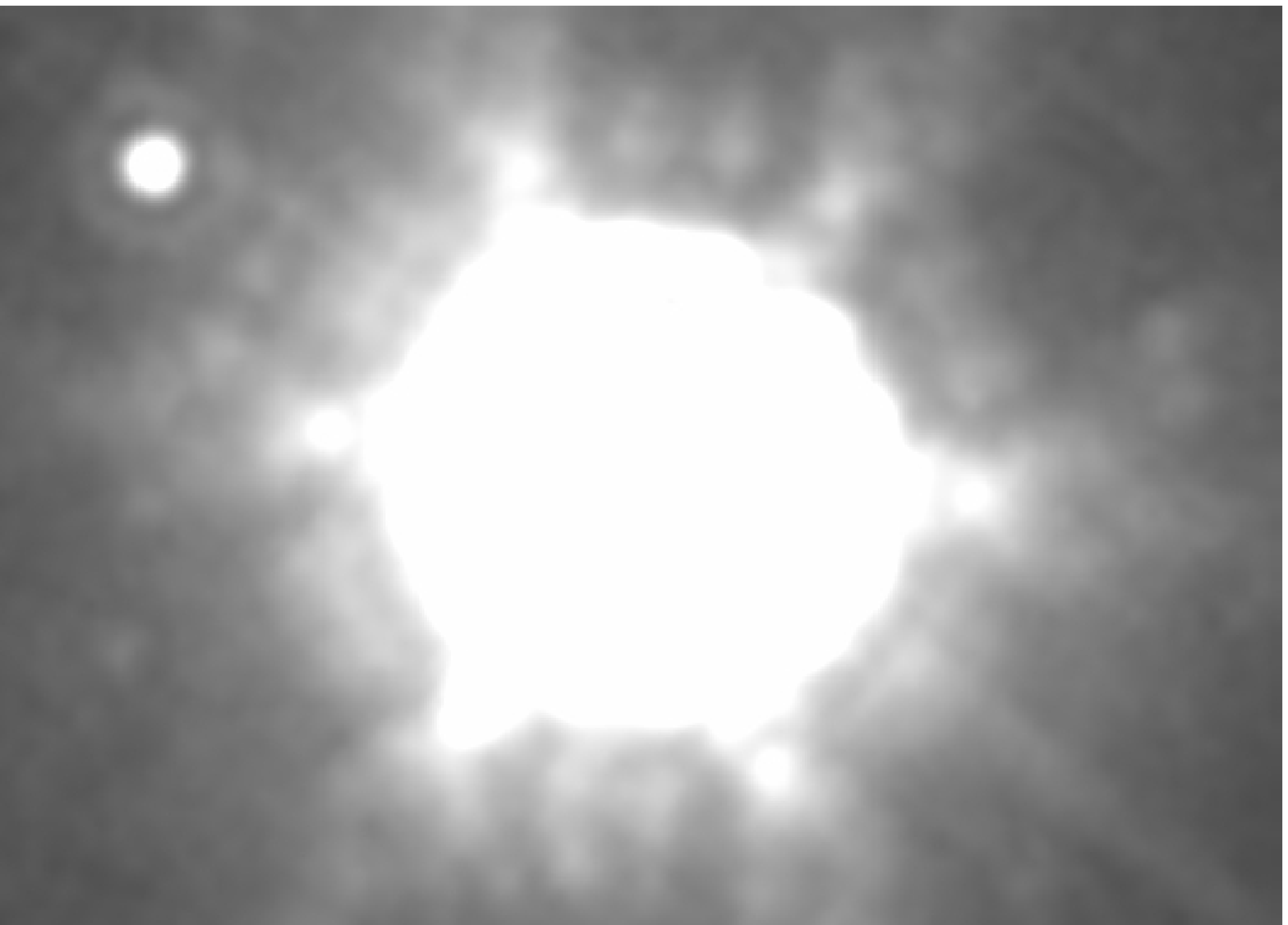} 
\includegraphics[angle=0,scale=.321]{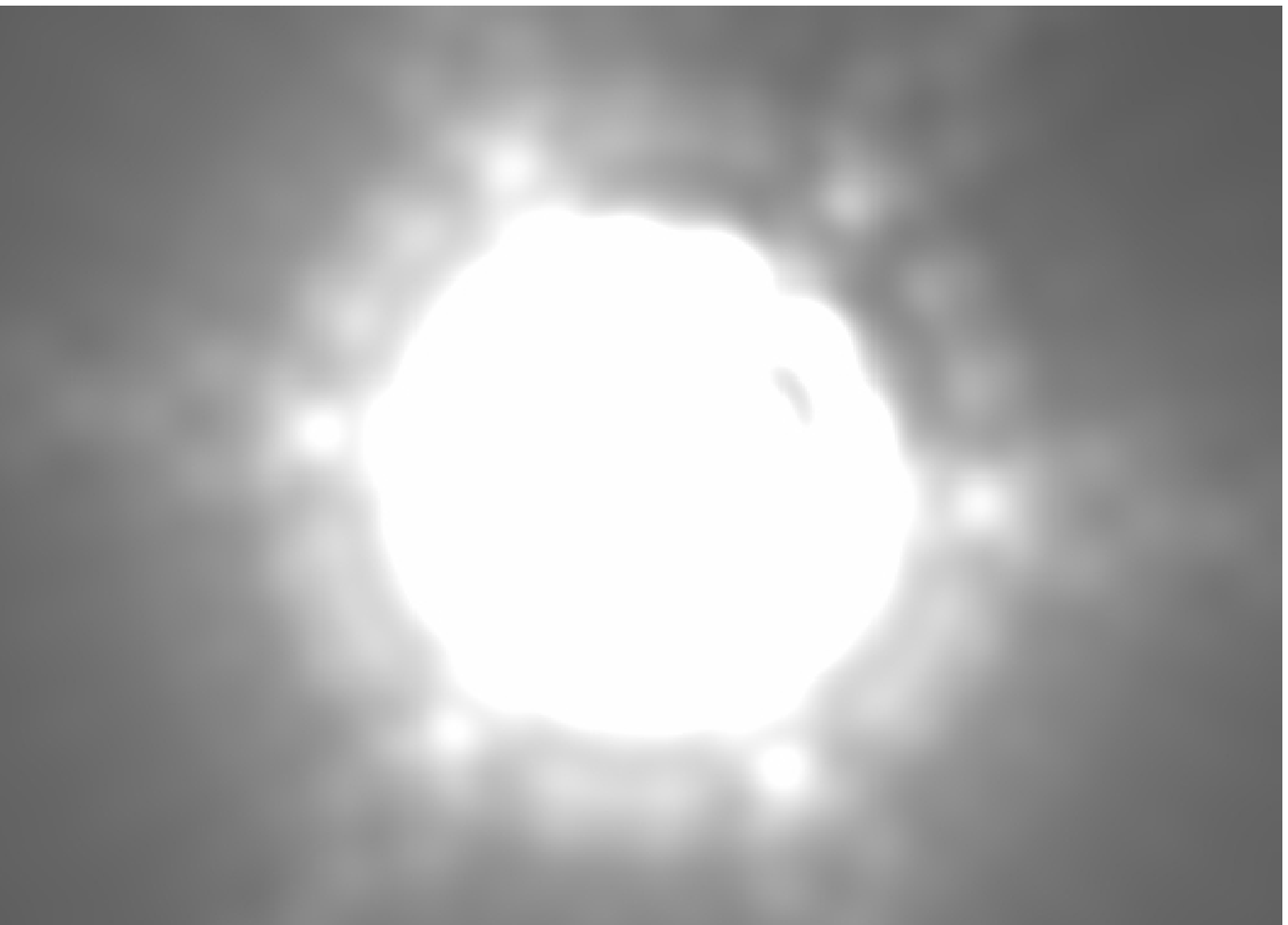} 
\includegraphics[angle=0,scale=.321]{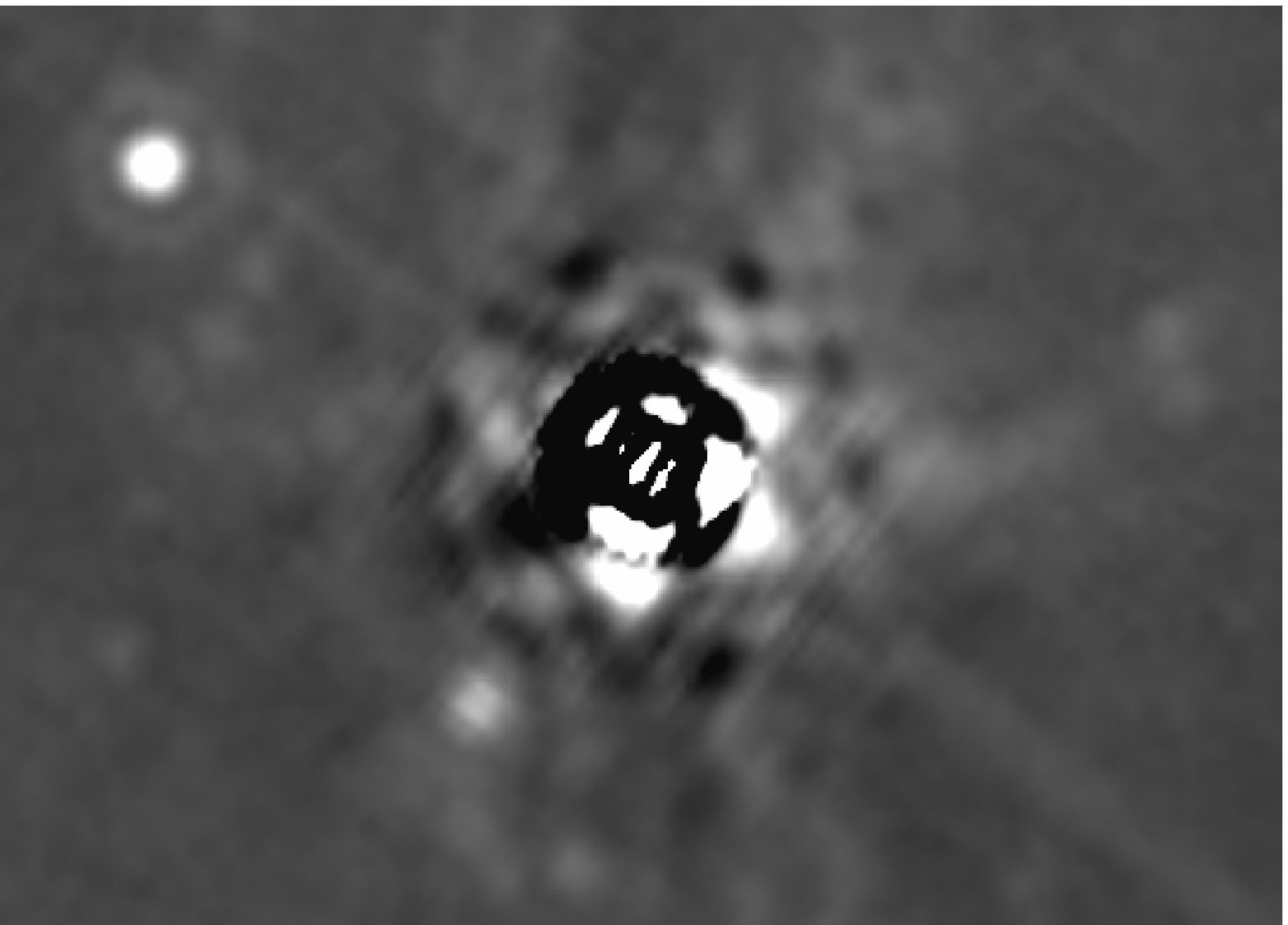} 

\includegraphics[angle=0,scale=.306]{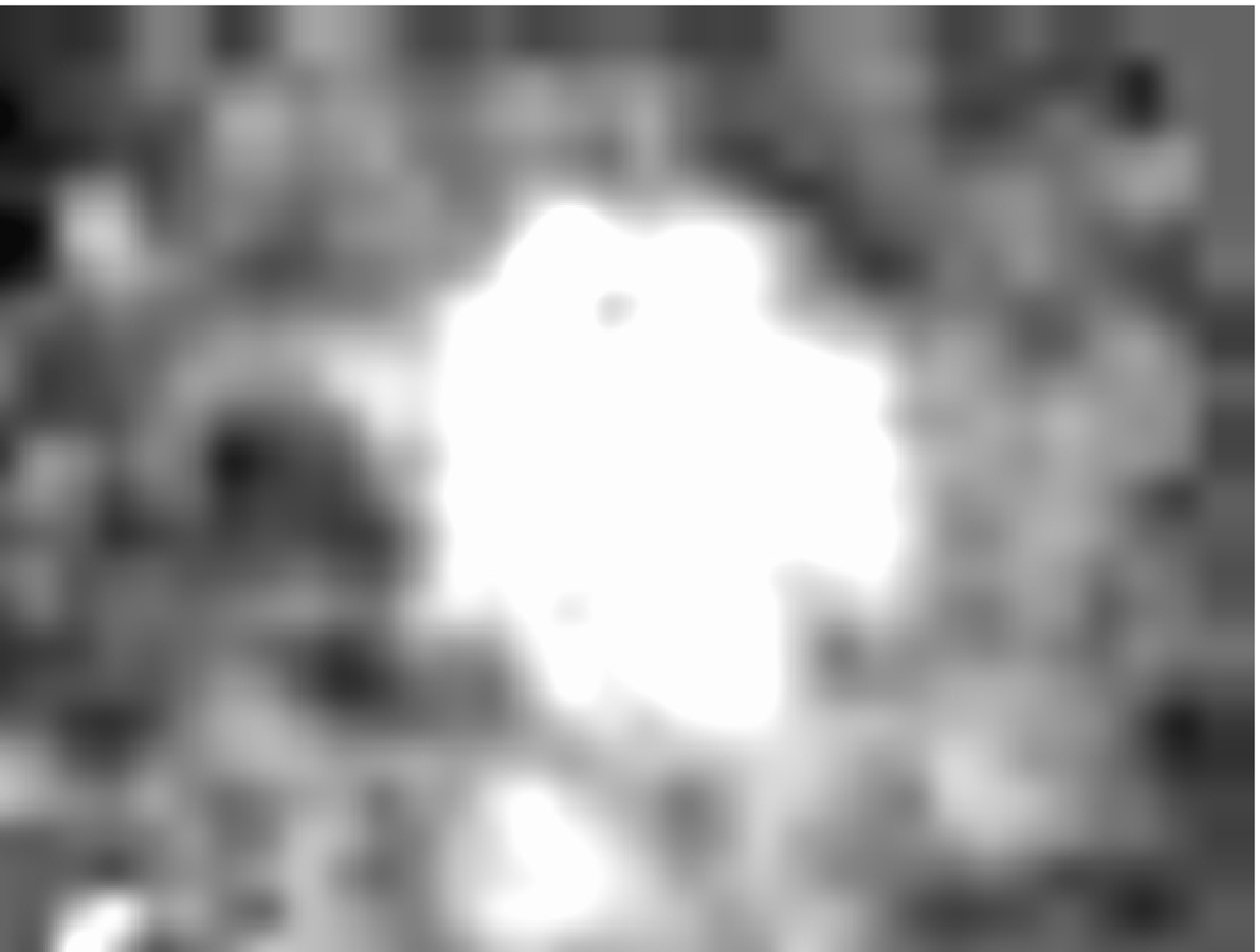} 
\includegraphics[angle=0,scale=.306]{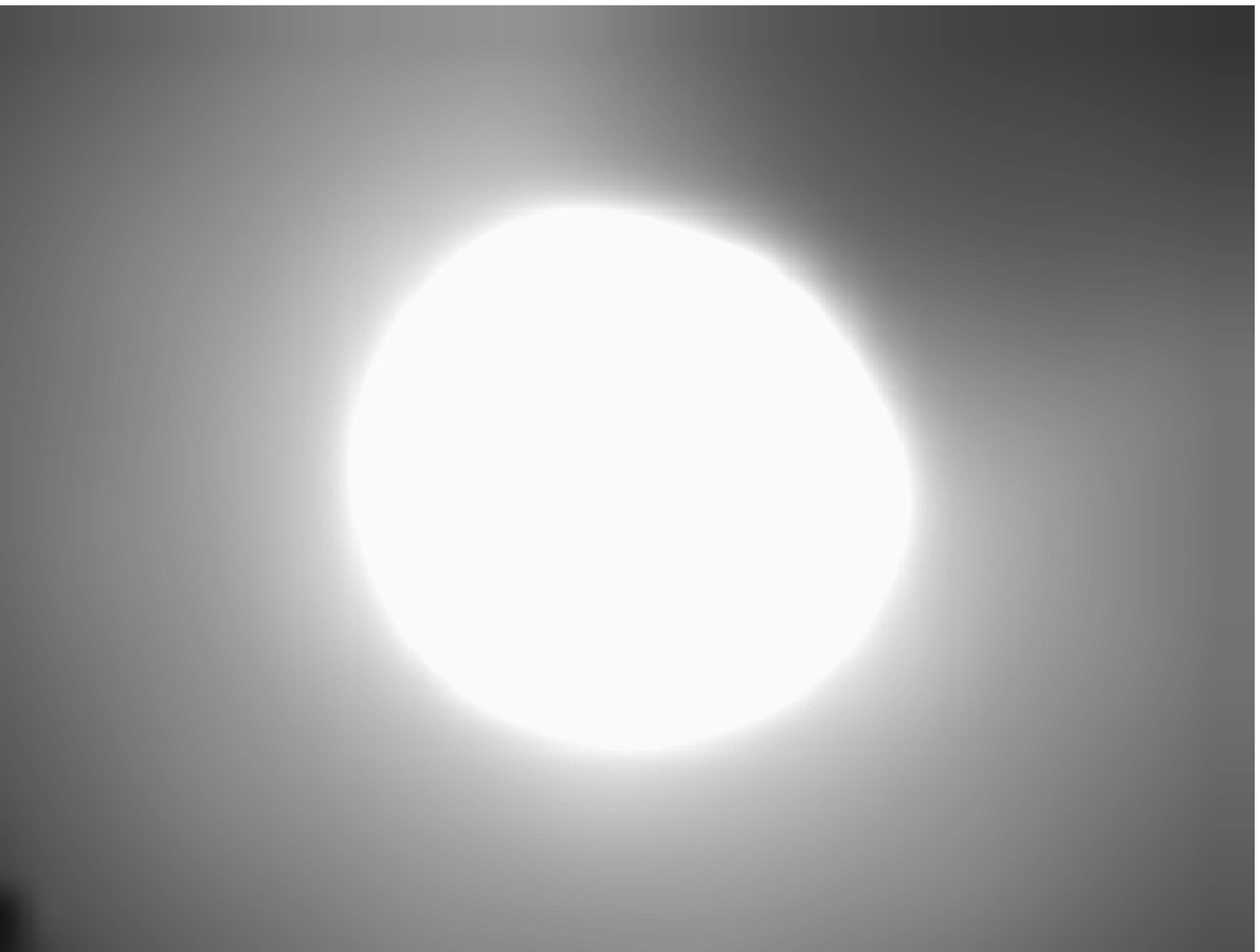} 
\includegraphics[angle=0,scale=.306]{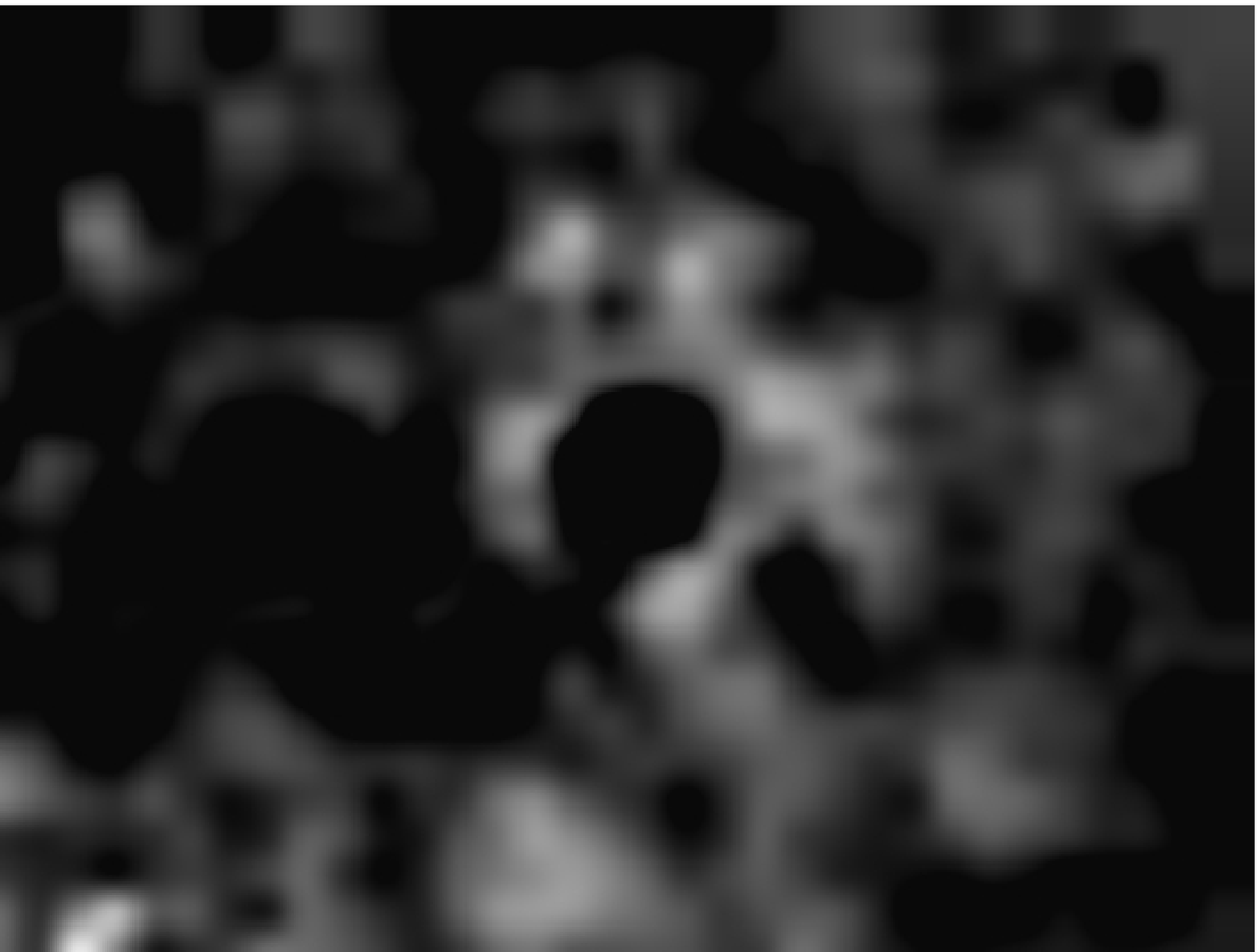} 
\caption{{\bf Top row:} \textit{Left panel:} Observed $24 ~\micron$ image. \textit{Center panel:} Model $24 ~\micron$ image including both stellar photosphere and bow shock. 
\textit{Right panel:} Model image subtracted from observed.
{\bf Bottom row:} \textit{Left panel:} Observed $70 ~\micron$ image. \textit{Center panel:} Model $70 ~\micron$ image including both stellar photosphere and bow shock. 
\textit{Right panel:} Model image subtracted from observed. The FOV is $\sim 2\farcm7 \times 2\farcm1$, N is up and E to the left.}
\label{fig:24-70model}
\end{figure*}

\begin{figure}[ht]
\figurenum{8}
\begin{center}
\includegraphics[angle=0,scale=.68]{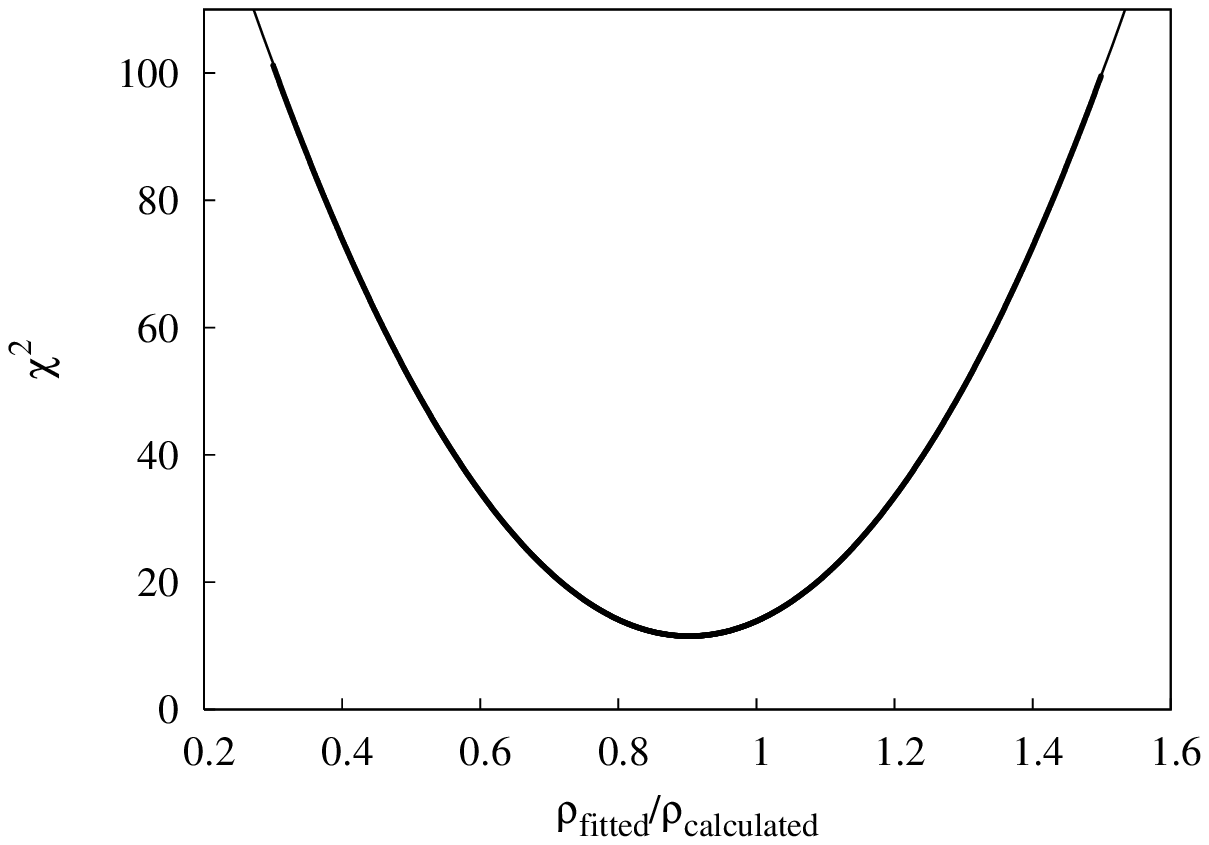}
\caption{The final ISM density was determined from the best fitting surface brightness image. This plot shows the $\chi^2$ of the fits of the model to the observed image, where
$\rho_{\rm calculated}$ is the initial guess from Figure \ref{fig:chi} {\it right panel} and $\rho_{\rm fitted}$ is fitted density using surface brightness values.}
\label{fig:bestrho}
\end{center}
\end{figure}

The $\chi^2$ of the deviations of the model from the observed image were calculated. 
We were able to constrain the rotation angle of the model and the relative velocity of 
the cloud to the star. The $\chi^2$ values in the $\varphi$ vs.\ $v_{\rm rel}$ phase 
space are shown in Figure \ref{fig:bestrot} ({\it left panel}). The small values at 
large rotation angles are artifacts due to the masking. The best-fit rotation angle is 
at $\varphi=-4^{\circ}$ to our initial guess, which means that the direction of motion 
is $143^{\circ}$ (CCW) of N. This is just $21^{\circ}$ from the proper motion direction. 
The ISM velocity predicted from vectorial velocity summation to fit this angle is 
$24 ~{\rm km~s}^{-1}$, which is close to the ISM velocity value calculated by 
\cite{lallement95}. The tangential velocity direction of the ISM from the summation is 
$\sim 47^{\circ}$ CW of N, which is pointing only $4\fdg5$ south from the galactic 
plane.

The $v_{\rm rel}$ parameter and its error are calculated by fitting a Gaussian to the 
phase space values at $\varphi=-4^{\circ}$ (Figure \ref{fig:bestrot}, {\it right panel}). 
One $\sigma$ errors are given by values at $\Delta\chi^2=1$. The fits give 
$v_{\rm rel}= 35.8 \pm 4.0 ~{\rm km~s}^{-1}$. Figure \ref{fig:chi} shows that if 
$v_{\rm rel}$ is constrained, then we can also determine the density of the cloud from 
simple SED modeling. The vertical cut of Figure \ref{fig:chi} ({\it left panel}) at 
$v_{\rm rel}= 36 ~{\rm km~s}^{-1}$ is shown in the {\it right panel} of the 
same Figure. 
$\rho_{\rm dust} = 6.43 \times 10^{-26} ~{\rm g~cm}^{-3}$  is derived from this fit, 
which gives an original ISM density of $6.43 \times 10^{-24}~{\rm g~cm}^{-3}$ assuming
the usual 1:100 dust to gas mass ratio. This $\rho_{\rm dust}$ is an upper estimate of 
the actual value, since the model computes what density would be needed to give the 
observed brightness using a $r_{\rm ext}$ radius sphere. Since the line-of-sight 
distribution of the dust is not cut off at $r_{\rm ext}$, we used 
$\rho_{\rm dust} = 6.43 \times 10^{-26}~{\rm g~cm}^{-3}$ as an initial guess; a range 
of density values was explored with model images. 

\begin{figure}[ht]
\figurenum{9}
\begin{center}
\includegraphics[angle=0,scale=.117]{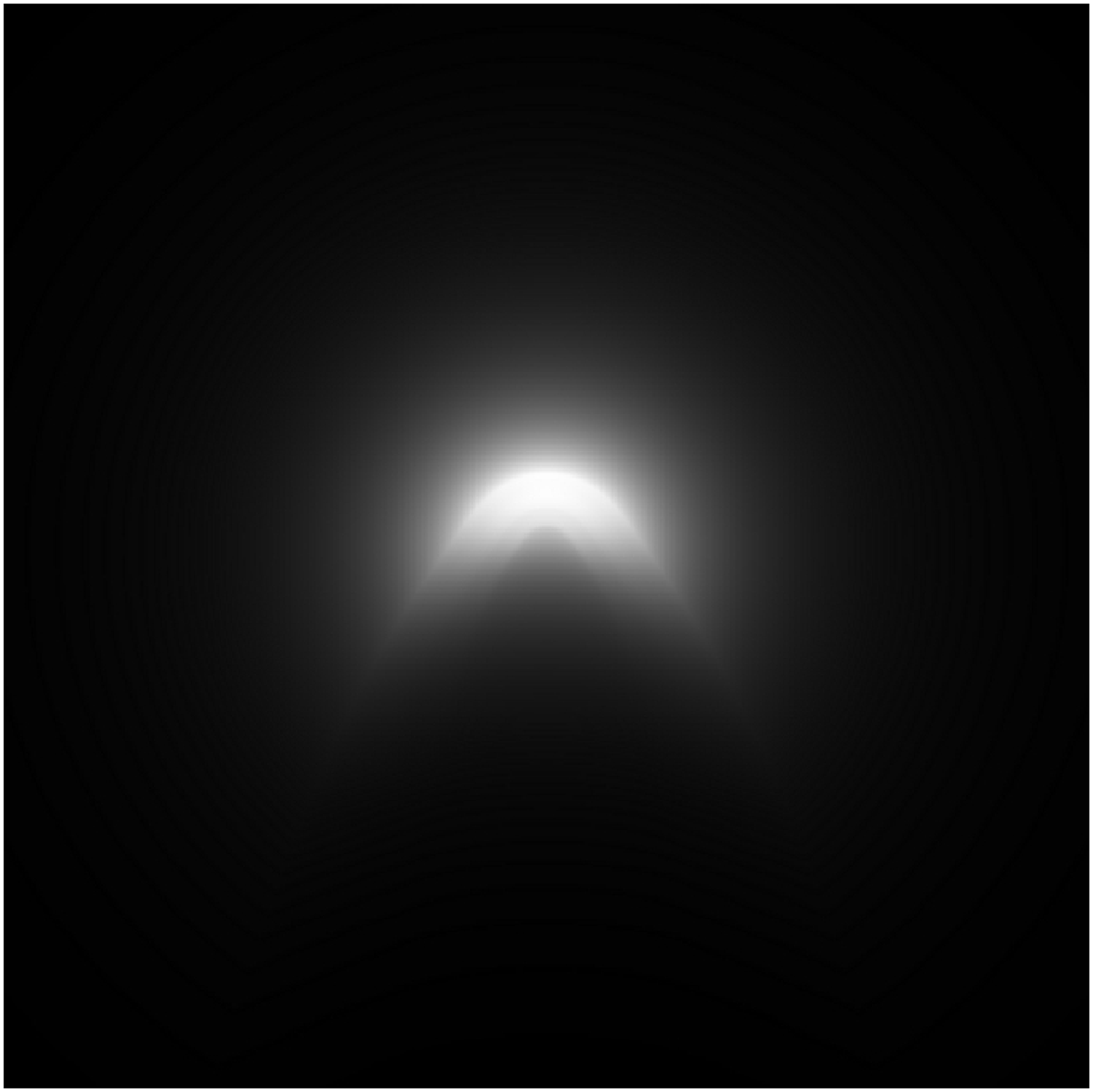}
\includegraphics[angle=0,scale=.117]{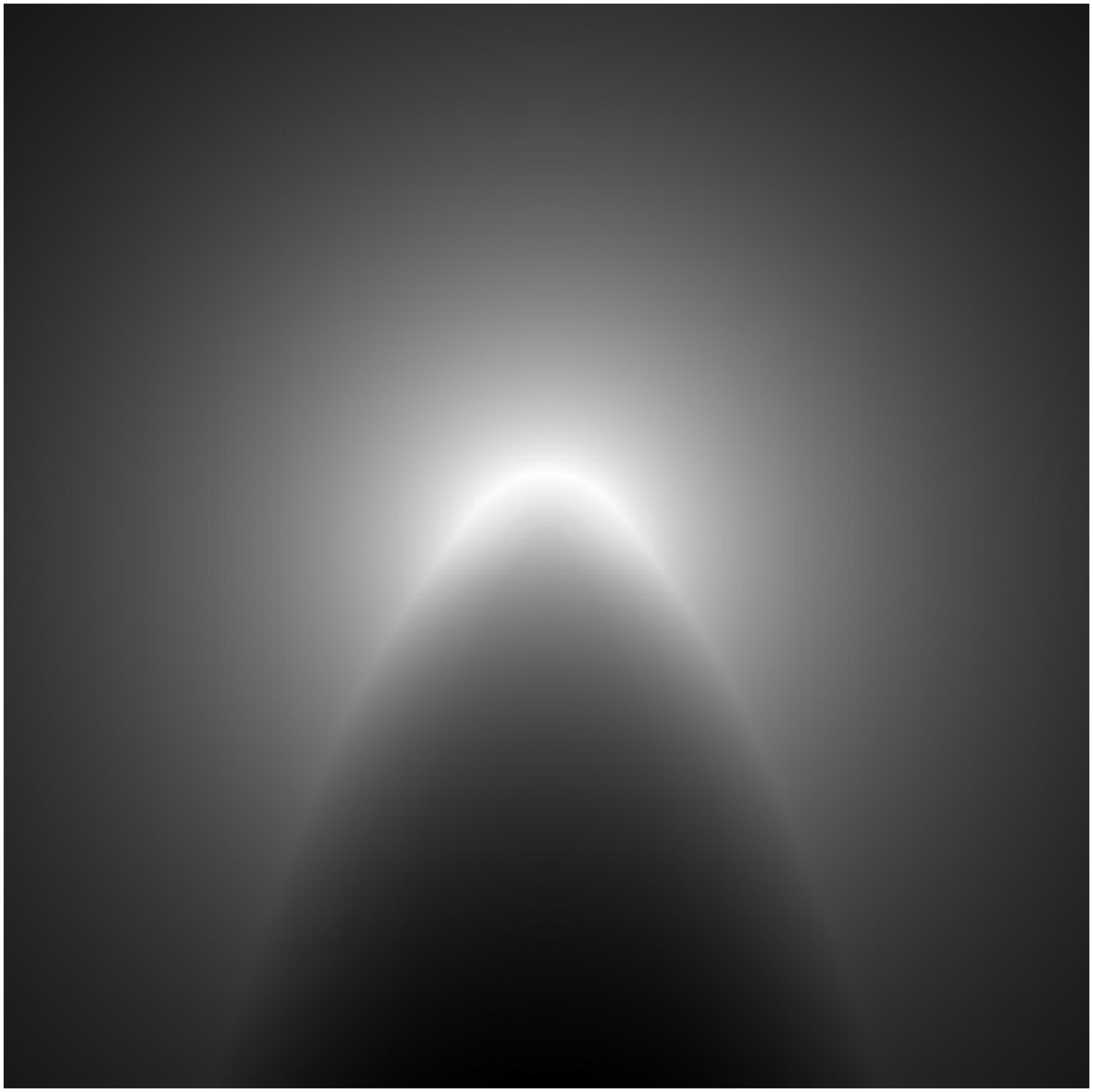}
\includegraphics[angle=0,scale=.117]{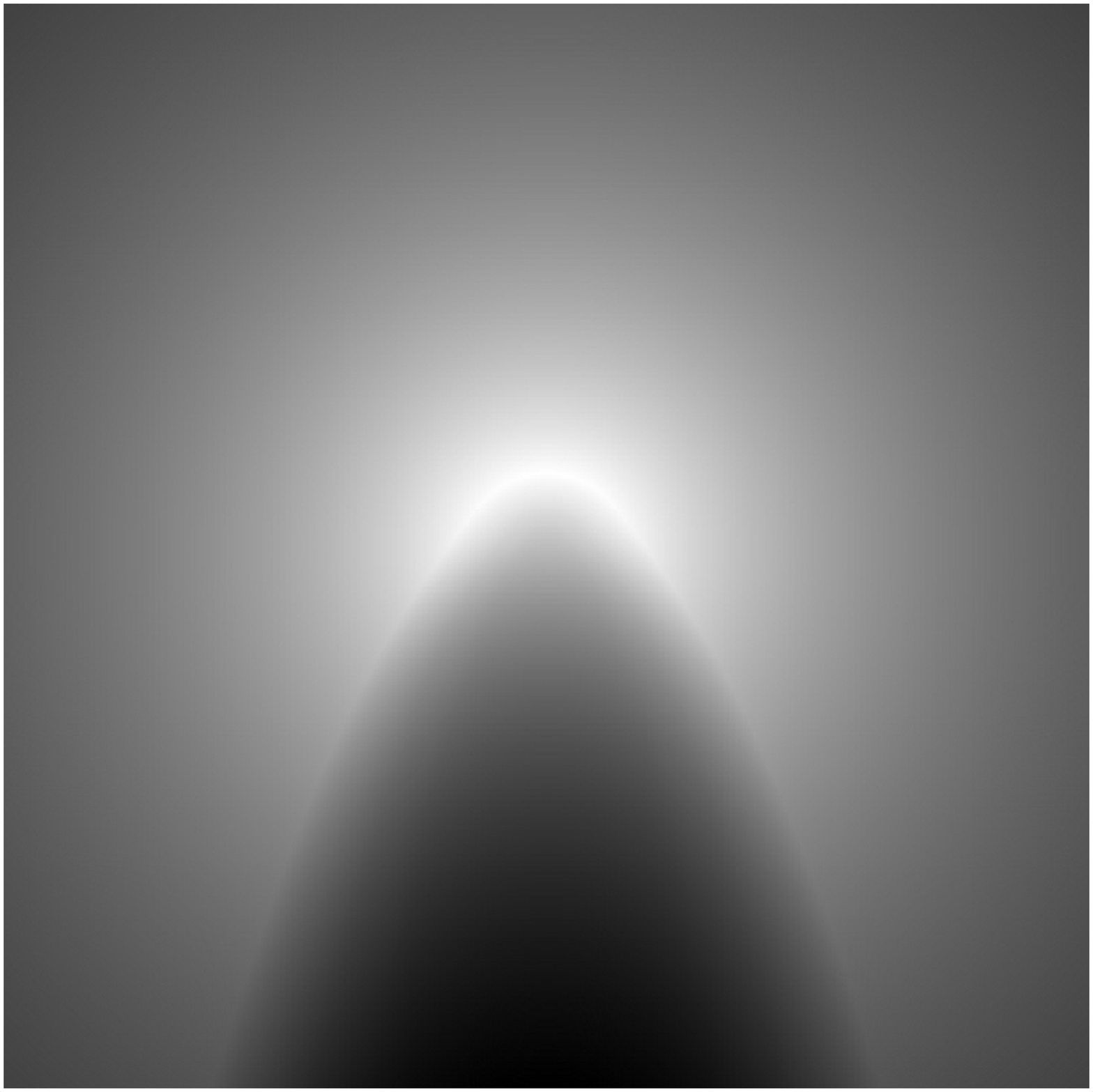}

\includegraphics[angle=0,scale=.2658]{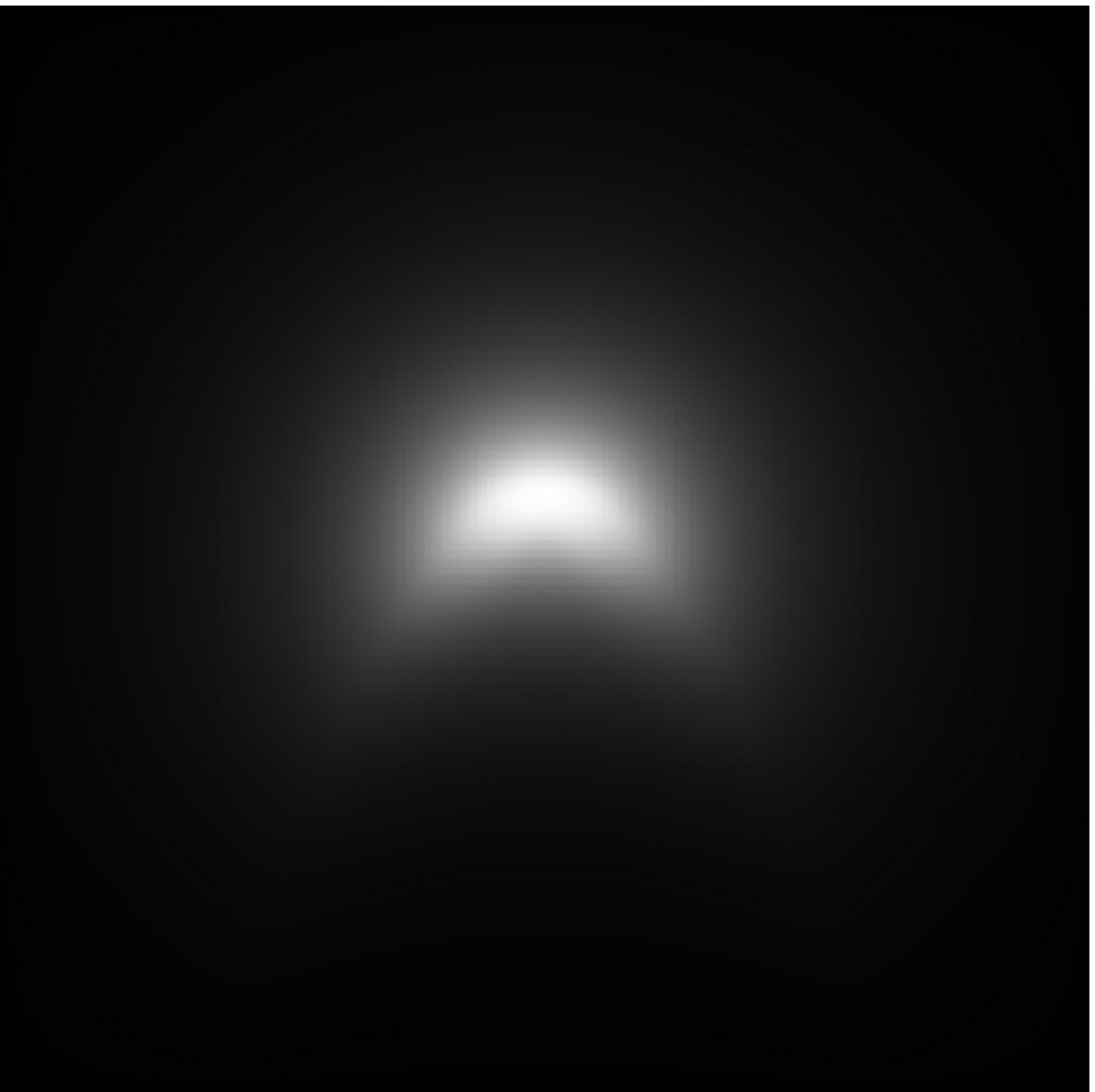}
\includegraphics[angle=0,scale=2.812]{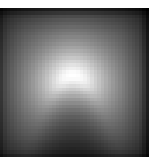}
\caption{\textit{Top panels:} Calculated high resolution surface brightnesses for $24$, $70$ and $160 ~\micron$, respectively. \textit{Bottom panels:} 
The $24$ and $70 ~\micron$ image with MIPS resolution, convolved with STinyTim PSFs. The images are not rotated to the same angle as the observed bow shock.}
\label{fig:surf}
\end{center}
\end{figure}

\begin{figure}[ht]
\figurenum{10}
\begin{center}
\includegraphics[angle=0,scale=.46]{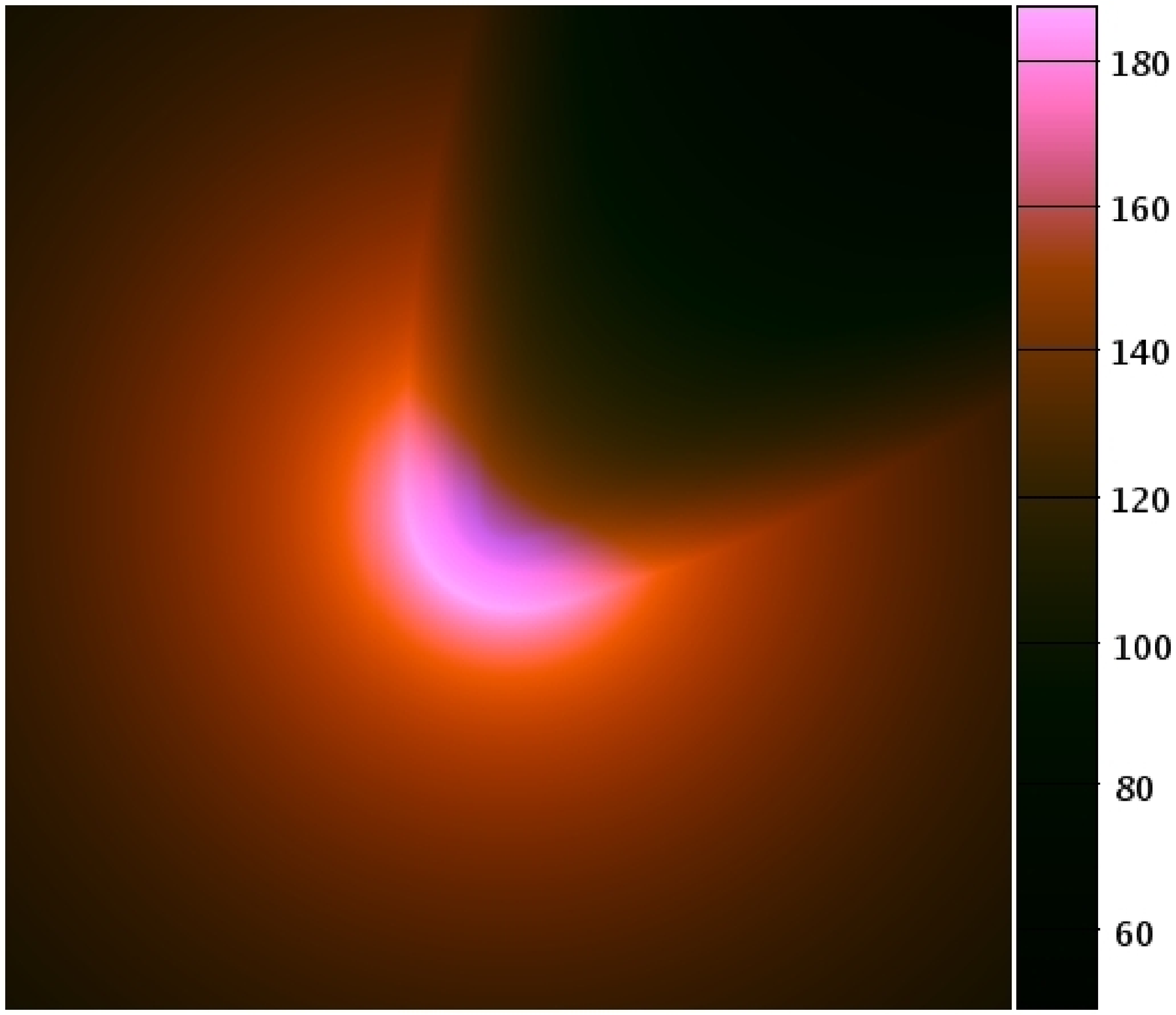}
\caption{Image (online version in color) of the bow shock generated by the model computations. The image's FOV is $2\farcm41 \times 2\farcm41$. The colorscale shows 
the integrated surface brightness temperature of the bow shock (and not the radial temperature gradient of the grains) in Kelvins.}
\label{fig:color}
\end{center}
\end{figure}

Since the surface brightness scales with the density, only one image had to be computed, 
which could be scaled afterwards with a constant factor. The resultant $\chi^2$ 
distribution is shown in Figure \ref{fig:bestrho}. The calculated best fitting ISM 
density is $5.8 \pm 0.4 \times 10^{-24}~{\rm g~cm}^{-3}$, assuming the average 1:100 
dust to gas mass ratio. The error was calculated at $\Delta\chi^2=1$. This density 
($n\sim 3.5 ~{\rm atoms~cm}^{-3}$) is only moderately higher than the average galactic ISM 
density ($\sim 1 ~{\rm atom~cm}^{-3}$). The calculated surface brightness images for 
the three MIPS wavelengths are shown in Figure \ref{fig:surf}. The closest stagnation 
point is for the $0.005 ~\micron$ grains at $64 ~{\rm AU}$, while the furthest is at 
$227 ~{\rm AU}$ for $0.056 ~\micron$ grains. The temperature coded image in Figure 
\ref{fig:color} shows the surface brightness temperature of the bow shock (i.e.\ the 
temperature of a black body, that would give the same surface brightness in the MIPS 
wavelengths as observed).  Table \ref{tab:par} shows good agreement between the model 
and the measured values.

\begin{figure}[ht]
\figurenum{12}
\begin{center}
\includegraphics[angle=0,scale=.68]{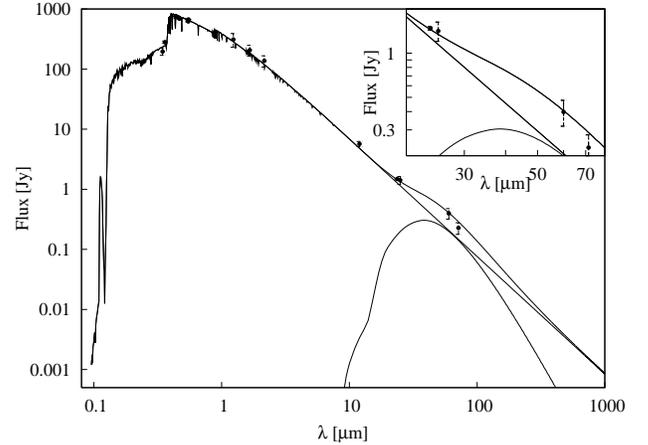}
\caption{The best fit SED. The window in the upper-right corner is a magnified part of the SED between $20$ and $80 ~\micron$. The plotted fluxes are $24$ and
$70 ~\micron$ MIPS and $25$ and $60 ~\micron$ IRAS, with errorbars. The $9.7 ~\micron$ silicate feature in the model SED of the ISM cloud is very faint and
on a bright continuum. The flux from the G dwarf has been subtracted from the $24$ and $70 ~\micron$ MIPS observations.}
\label{fig:sed}
\end{center}
\end{figure}

The original observed images at $24 ~\micron$ and $70 ~\micron $ were compared to the 
model. We generated model images with high resolution that included the bow shock and 
the central star with its photospheric brightness value at the central pixel. We 
convolved these images with a $1.8$ native pixel boxcar smoothed STinyTim PSF
\citep[see][]{engelbracht07}. These images were subtracted from the 
observed ones (Figure \ref{fig:24-70model}). The residuals are small and generally 
consistent with the expected noise. Finally, the best fitting SED of the system 
($r_{\rm ext}=1366 ~{\rm AU}$) is plotted in Figure \ref{fig:sed}.
The total mass of the  dust inside the $r_{\rm ext}=1366 ~{\rm AU}$ radius is 
$M_{\rm dust} = 1.706\times10^{24}~{\rm g}$ ($0.023 ~M_{\rm Moon}$).

\section{Discussion}
\subsection{Bow Shock Model Results}

Our model gives a consistent explanation of the total infrared excess and the 
surface brightness distribution of the bow shock structure at $\delta$ Velorum. The 
question still remains how common this phenomenon is among the previously identified 
infrared-excess stars. Is it possible that many of the infrared excesses found around 
early-type stars result from the emission of the ambient ISM cloud? The majority of 
infrared excess stars are distant and cannot be resolved, so we cannot answer for sure. 
However the excess at $\delta$ Velorum is relatively warm between $24$ and $70 ~\micron$ 
(${\rm F}(24) \sim 0.17 ~{\rm Jy}$, ${\rm F}(70) \sim 0.14 ~{\rm Jy}$), and such 
behavior may provide an indication of ISM emission. This possibility will be analyzed in 
a forthcoming paper. Another test would be to search for ISM spectral features. The ISM 
$9.7 ~\micron$ silicate feature of the dust grains would have a total flux of 
$\sim 1 ~{\rm mJy}$ for $\delta$ Velorum. Since the $\sim 1 ~{\rm mJy}$ flux would originate 
from an extended region and not a point source that could fit in the slit of IRS, it would be nearly
impossible to detect with {\it Spitzer}. Only a faint hint of the excess is visible in the
$8 ~\micron$ IRAC images, consistent with the small output predicted by our model.

\subsection{ISM Interactions}

To produce a bow shock feature as seen around $\delta$ Velorum, the star needs 
to be luminous, have a rather large relative velocity with respect to
the interacting ISM, and be passing through an ISM cloud. A relative velocity of 
$\sim 36~{\rm km~s}^{-1}$ is not 
necessarily uncommon, since the ISM in the solar neighborhood has a space velocity of 
$\sim 26~{\rm km~s}^{-1}$ \citep{lallement95} and stars typically move with similar 
speeds. If the ISM encountering the star is not dense enough the resulting excess will 
be too faint to be detected. The Sun and its close ($\sim 100~{\rm pc}$) surrounding are 
sitting in the Local Bubble 
\citep[${\rm n(HI)}< 0.24 ~{\rm cm}^{-3}$, ${\rm T} \approx 7500 ~{\rm K}$,][]{lallement98,jenkins02}.
This cavity generally lacks cold and neutral gas up to $\sim 100 ~{\rm pc}$. The density 
we calculated at $\delta$ Velorum is $\sim 15$ times higher than the average value 
inside the Local Bubble. Observations over the past thirty years have shown that this 
void is not completely deficient of material, but contains filaments and cold clouds 
\citep{wennmacher92,herbstmeier98,jenkins02,meyer06}. \cite{talbot77} calculated that an 
average galactic disk star of solar age has probably passed through about 135 clouds of
${\rm n(HI)} \ge 10^2 ~{\rm cm}^{-3}$ and about 16 clouds with 
${\rm n(HI)} \ge 10^3 ~{\rm cm}^{-3}$. Thus the scenario that we propose for $\delta$ 
Velorum is plausible.

\subsection{Implications for Diffusion/Accretion Model of $\lambda$ Bo\"otis Phenomenon}

\begin{figure}[ht]
\figurenum{13}
\begin{center}
\includegraphics[angle=0,scale=.68]{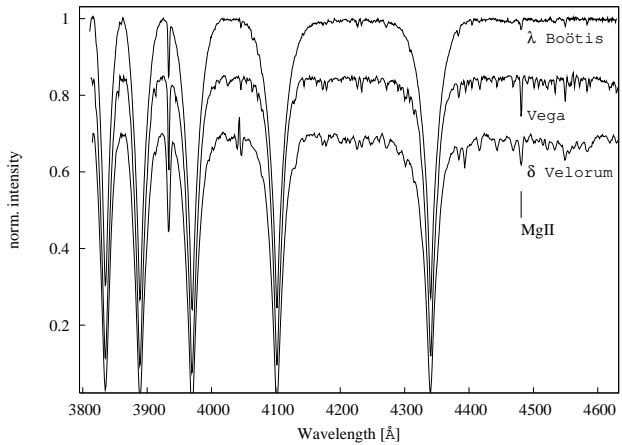}
\caption{The spectra of $\delta$ Velorum (bottom line), Vega (middle line) and $\lambda$ Bo\"otis (top line).}
\label{fig:spectr}
\end{center}
\end{figure}

\cite{holweger99} list $\delta$ Velorum as a simple A star, not a $\lambda$ 
Bo\"otis one. We downloaded spectra of the star from  the Appalachian State University 
Nstars Spectra Project \citep{gray06}. The spectra of $\delta$ Velorum, $\lambda$ 
Bo\"otis (prototype of its group) and Vega (an MK A0 standard) are plotted in Figure 
\ref{fig:spectr}. The metallic lines are generally strong for $\delta$ Velorum. One of 
the most distinctive characteristics of $\lambda$ Bo\"otis stars is the absence or 
extreme weakness of the MgII lines at $4481$ \AA{} \citep{gray88}. Although the MgII 
line seems to be weaker than expected for an A0 spectral type star, it still shows high 
abundance, which confirms that $\delta$ Velorum is not a $\lambda$ Bo\"otis type star 
(Christopher J. Corbally, private communication). The overall metallicity ratio for 
$\delta$ Velorum is $\left[{\rm M}/{\rm H}\right]=-0.33$, while for $\lambda$ Bo\"otis 
it is $\left[{\rm M}/{\rm H}\right]=-1.86$ \citep{gray06}. The G star's contribution to 
the total abundance in the spectrum is negligible, because of its relative faintness. We 
used spectra from the NStars web site to synthesize a A1V/A5V binary composite spectrum and 
found only minor differences from the A1V spectrum alone. Thus, the assigned metallicity 
should be valid.

These results show that at $\delta$ Velorum, where we do see the ISM interacting with a star, 
there is no sign of the $\lambda$ Bo\"otis phenomenon or just a very mild effect. 
\cite{turcotte93} modeled that an accretion rate of $\sim 10^{-14} ~M_{\odot}~{\rm yr}^{-1}$ is
necessary for a $T_{\rm eff}=8000~{\rm K}$ main sequence star to show the spectroscopic 
characteristics of the phenomenon. The $\lambda$ Bo\"otis abundance pattern starts to 
show at $10^{-15} ~M_{\odot}~{\rm yr}^{-1}$ and ceases at 
$10^{-12}~M_{\odot}~{\rm yr}^{-1}$. To reach an ISM accretion of 
$10^{-15}~M_{\odot}~{\rm yr}^{-1}$ a collecting area of $2 ~{\rm AU}$ radius would be 
needed with our modeled ISM density and velocity.
For an accretion of $10^{-14}$, $10^{-13}$ and $10^{-12} ~M_{\odot}~{\rm yr}^{-1}$ 
collecting areas of $6.5$, $20$ and $65 ~{\rm AU}$ radii are needed, respectively. 

With the accretion theory of \cite{bondi44}, we get an accretion rate of 
$6.15 \times 10^{-15}~M_{\odot}~{\rm yr}^{-1}$ for $\delta$ Velorum. Thus, the accretion rate for 
this star is probably not high enough to show a perfect $\lambda$ Bo\"otis spectrum, but should be high 
enough for it to show some effects of accretion. This star is an exciting 
testbed for the diffusion/accretion model of the $\lambda$ Bo\"otis phenomenon.

\section{Summary}

We observe a bow shock generated by photon pressure as $\delta$ Velorum moves 
through an interstellar cloud. Although this star was thought to have a debris disk, its 
infrared excess appears to arise at least in large part from this bow shock. We present 
a physical model to explain the bow shock. Our calculations reproduce the observed 
surface brightness of the object and give the physical parameters of the cloud. We 
determined the density of the surrounding ISM to be 
$5.8 \pm 0.4 \times 10^{-24} ~{\rm g~cm}^{-3}$. This corresponds to a number density of
$n \approx 3.5 ~{\rm atoms ~cm}^{-3}$, which means a $\sim 15$ times overdensity relative to 
the average Local Bubble value. The cloud and the star have a relative velocity of 
$35.8 \pm 4.0 ~{\rm km~s}^{-1}$. The velocity
of the ISM in the vicinity of $\delta$ Velorum we derived is consistent with LIC 
velocity measurements by \cite{lallement95}. Our best-fit parameters and measured fluxes 
are summarized in Table \ref{tab:par}.

\cite{holweger99} found that $\delta$ Velorum is not a $\lambda$ Bo\"otis star. The
measurements from the Nstars Spectra Project also confirm this. Details regarding the 
diffusion/accretion time scales for a complex stellar system remain to be elaborated. 
Nevertheless, our {\it Spitzer} observations of $\delta$ Velorum provide an interesting 
testbed and challenge to the ISM diffusion/accretion theory for the $\lambda$ Bo\"otis 
phenomenon.

\acknowledgments
Based on observations with \textit{Spit\-zer Space Telescope}, which is 
operated by the Jet Propulsion Laboratory, California Institute of Technology under NASA 
contract 1407. Support for this work was provided by NASA through Contract Number 1255094 
issued by JPL/Caltech. This research made use of the SIMBAD database, operated at CDS, 
Strasbourg, France. We would like to thank the help of A.\ Skemer in the error analysis 
of the model. We have benefited from the helpful discussions with C.\ Corbally and 
D.\ Apai. The analysis of IRAC data by J.\ Carson helped to confirm our results.

{\it Facilities:} \facility{{\it Spitzer} (MIPS)}

\clearpage

\end{document}